\documentclass[%
 reprint,
superscriptaddress,
 amsmath,amssymb,
 aps,
]{revtex4-2}

\usepackage{graphicx}
\usepackage{dcolumn}
\usepackage{bm}

\begin{document}


\title{Coherent synchrotron radiation instability in low-emittance electron storage rings}

\author{Sara Dastan}
\affiliation{%
 Elettra Sincrotrone Trieste, Basovizza, Italy 
}%

\author{Demin Zhou}
\email{dmzhou@post.kek.jp}
\affiliation{%
 KEK, 1-1 Oho, Tsukuba 305-0801, Japan 
}%
\affiliation{
 The Graduate University for Advanced Studies, SOKENDAI
}%

\author{Takuya Ishibashi}
\affiliation{%
 KEK, 1-1 Oho, Tsukuba 305-0801, Japan 
}%
\affiliation{
 The Graduate University for Advanced Studies, SOKENDAI
}%

\author{Emanuel Karantzoulis}
\affiliation{%
 Elettra Sincrotrone Trieste, Basovizza, Italy 
}%

\author{Simone Di Mitri}
\affiliation{%
 Elettra Sincrotrone Trieste, Basovizza, Italy 
}%

\author{Ryan Lindberg} 
\affiliation{Argonne National Laboratory, Argonne, Illinois 60439, USA}


\date{\today}

\begin{abstract}
Longitudinal impedances at high frequencies, which extend far beyond the width of the beam spectrum, can pose a threat to the performance of modern low-emittance electron storage rings, as they can establish a relatively low threshold for microwave instability. In such rings, coherent synchrotron radiation (CSR) emerges as a prominent contributor to these high-frequency impedances. This paper undertakes a systematic investigation into the effects of CSR on electron rings, utilizing Elettra 2.0, a ring of fourth-generation light sources, and the SuperKEKB low-energy ring, a ring of e+e- circular colliders, as illustrative examples. Our work revisits theories of microwave instability driven by CSR impedance, extending the analysis to encompass other high-frequency impedances such as resistive wall and coherent wiggler radiation. Through instability analysis and numerical simulations conducted on the two aforementioned rings, the study explored the impact of high-frequency impedances and their interactions with broadband impedances from discontinuities in vacuum chambers.
\end{abstract}

\maketitle


\section{\label{sec:intro}Introduction}

Coherent synchrotron radiation (CSR) generated by charged particles moving along a curved orbit, primarily induced by bending or wiggler magnets, plays a crucial role in triggering beam instabilities in both electron linacs, particularly free electron lasers, and electron storage rings (for a comprehensive review, see Secs.~2.4.16 and~3.1.3 of~\cite{chao2023handbook}). Although the CSR field that affects a charged beam is often treated as a type of wakefield~\cite{zotter1998impedances, chao1993physics}, its overtaking nature~\cite{derbenev1995microbunch} sets it apart. Here, ``overtaking'' denotes the scenario in which the radiation fields of the trailing particles traveling along the chords catch up and influence the leading particles as a charged beam moves along a curved orbit. Another form of overtaking wakefield arises from the space charge force generated by the Coulomb fields of charged particles when the beam is moving along a straight orbit but with relativistic velocity $\beta=v/c<1$ ($c$ is the speed of light in vacuum). The existence of metal walls slows the fields, causing the CSR fields to include a trailing component~\cite{agoh2009steady}. In other words, in the CSR problem, the leading particles can also influence the trailing particles in a charged beam. This is more justified in the most general case of $\beta<1$ with the presence of resistive and non-smooth walls. Then, the CSR problem involves a combination of CSR, space charge, and conventional wakefields due to the non-smooth chamber (for a relevant example, see~\cite{billinghurst2015observation}), and separating these three components is not trivial. 

In recent decades, extensive research has been done on modeling the CSR impedance, along with the development of theories, simulations, and measurements of the CSR instability in electron storage rings. See references in Sec.~2.4.16 of~\cite{chao2023handbook} and recent publications such as~\cite{agoh2023transient, blednykh2023microwave} for further details. In line with this trend, deliberate efforts have been made to induce the CSR instability intentionally in electron rings, with the aim of generating THz lights for various applications (for example, see~\cite{evain2019stable, brosi:ipac21-thxa02} and references therein). A convenient scaling law to check the instability threshold for the bunch current is given by~\cite{bane2010threshold}
\begin{equation}
    I_\text{th1}=
    \frac{4\pi(E/e)\eta\sigma_\delta^2\sigma_z^{1/3}}{Z_0\rho^{1/3}}
    S_\text{th1}
    \label{eq:IthCSR}
\end{equation}
with 
\begin{equation}
    S_\text{th1}\approx 0.5+0.12\Pi,
    \label{eq:Sth1}
\end{equation}
where $\Pi=\sigma_z\sqrt{\rho/h^3}$ indicating the shielding effects of parallel plates. Relevant quantities in the equations are: $e$ the electron charge, $Z_0$ the impedance of the free space, $E$ the beam energy, $\eta=\alpha_p-1/\gamma^2$ the slip factor ($\alpha_p$ is the momentum compaction factor and $\gamma$ is the relativistic factor), $\sigma_\delta$ the nominal energy spread, $\sigma_z$ the nominal bunch length, $\rho$ the bending radius in dipole magnets, and $2h$ the full height of the vacuum chamber in dipole magnets. The synchrotron tune $\nu_s$ in the original formulation of~\cite{bane2010threshold} is replaced by $\nu_s=c\eta\sigma_\delta/(\sigma_z\omega_0)$ with $c$ the speed of light and $\omega_0=2\pi c/C$ ($C$ is the circumference of the ring). Note that Eq.~(\ref{eq:IthCSR}) is valid only for positive $\eta$. For negative $\eta$, the scaling factor $S_\text{th1}$ must be modified~\cite{stupakov2002beam, blednykh2023microwave}. However, we omit the details in this paper, as our primary focus is on positive $\eta$, which is typically applicable to modern storage rings. For the free-space steady-state (FS-SS) CSR, the dependence of $I_\text{th1}$ on the sign of $\eta$ is discussed in~\cite{stupakov2002beam, blednykh2023microwave}. Equation~(\ref{eq:IthCSR}) was obtained by numerically solving the Vlasov-Fokker-Planck (VFP) equation and validated by experiments~\cite{brosi2019systematic} in which the CSR impedance dominates the microwave instability (MWI). In other words, Eq.~(\ref{eq:IthCSR}) solely determines the MWI threshold for an electron storage ring when the threshold for CSR instability is significantly lower than the MWI threshold induced by conventional wakes. Note that Eq.~(\ref{eq:IthCSR}) assumes that all dipole magnets have the same $\rho$ with their total length $2\pi\rho$ for the storage ring.

In fourth-generation light sources based on storage rings (commonly referred to as diffraction-limited light sources~\cite{yabashi2017next, raimondi2023toward}) and circular e+e- colliders such as SuperKEKB~\cite{SuperKEKBTDR}, the CSR instability is undesirable, as it poses a threat to their stable operations. During the design stage of these storage rings, it is essential to optimize the scaling parameter $\gamma\eta \sigma_\delta^2\sigma_z$~\cite{boussard1975observation}, with the objective of maximizing the MWI threshold. To assess the influence of CSR on the MWI threshold, as indicated by Eq.~(\ref{eq:IthCSR}), it is valuable to examine the scaling parameter $\gamma\eta\sigma_\delta^2(\sigma_z/\rho)^{1/3}$ and the shielding parameter $\Pi$. These parameters are contingent upon the configurations of the magnet and vacuum systems. However, the selection of $\rho$ and $h$ is not entirely unrestricted, as the technical and accelerator physics goals (such as achieving extremely small emittance) impose constraints on these parameters.

In this paper, we present a detailed investigation of the CSR instability in low-emittance electron storage rings. In such rings, hardware configurations are fixed, and detailed bottom-up impedance calculations are employed for subsequent investigations of beam instabilities. The CSR impedance is treated as a conventional impedance, but with special attention through instability analysis and a careful set-up of numerical simulations.

The paper is organized as follows. In Sec.~\ref{sec:csrmodels}, we briefly review alternative CSR impedance models for longitudinal single-bunch effects in electron storage rings. The theory of instability analysis with arbitrary impedance is discussed in Sec.~\ref{sec:instability}. Detailed investigations of CSR effects in two projects, Elettra 2.0 and SuperKEKB low energy ring (LER), are documented in Secs.~\ref{sec:csrElettra2} and~\ref{sec:csrSKBLER}, respectively. In Sec.~\ref{sec:summary}, we present our conclusions based on our findings.

\section{\label{sec:csrmodels}CSR impedance models for single-bunch effects in electron storage rings}

For low-emittance electron storage rings, the transverse beam size $\sigma_\perp$ is typically much smaller than the transverse coherence size $l_\perp \sim (\rho\sigma_z^2)^{1/3}$. This makes it sufficient to consider only 1D longitudinal models for the CSR impedance. So far, all CSR impedance models implicitly assume that the rigid-beam approximation~\cite{ChaoNotes2002, chao2020lectures} is valid. Before delving into detailed calculations, it is crucial to consider a few scaling parameters for quick evaluations:
\begin{itemize}
    \item The critical wavenumber of the synchrotron radiation: $k_c=3\gamma^3/(2\rho)$. The total longitudinal CSR impedance of a ring scales as $Z_\parallel(k)\propto (\rho k)^{1/3}$ when $k\ll k_c$ (see Eq.~(\ref{eq:csrFS})) and $Z_\parallel(k)\propto \gamma e^{-x}\sqrt{x}$ with $x=k/k_c$ when $k\gg k_c$. Coherent radiation occurs at $k\ll k_c$, and the interaction distance due to the CSR fields inside a bunch is $z\gg 1/k_c$.
    \item Wall shielding threshold: $k_w=\pi\sqrt{\rho/(2h)^3}$ (here, we quote Eq.~(2) from~\cite{agoh2009steady}, although the coefficient is defined differently by other authors). Suppose that the beam is moving along a curved orbit in the horizontal plane. The metal walls, symmetrically placed parallel to the orbit plane with distance $2h$, provide a strong shielding for the CSR. For a bunch length $\sigma_z\ll 1/k_w$, wall shielding is not critical for evaluating beam dynamics; however, for $\sigma_z\geq 1/k_w$, wall shielding becomes significant. This is evident from Eq.~(\ref{eq:Sth1}), where it can be observed that $S_\text{th1}\approx 0.5+0.11\sigma_zk_w$.
    \item Radiation formation length: $l_f=(24\rho^2\sigma_z)^{1/3}$. This parameter is compared to the magnet length $l_b$ to assess the impact of the transient CSR that occurs at the entrance and exit of the magnet. For long magnets where $l_b\gtrsim l_f$, transient effects are negligible; while for $l_b<l_f$, transient effects become significant.
    \item Catch-up distance: $l_c=2\sqrt{2\rho w}$, where $w$ is the distance from the orbit to the side walls symmetrically placed perpendicular to the orbit plane (see Fig.~6 of~\cite{zhou2011calculation}). The outer wall reflects the CSR fields that influence the trailing particles of the bunch; while the inner wall shields the overtaking CSR fields that affect the leading particles.
    \item Slippage length: $l_s=\eta\sigma_\delta C$. This parameter quantifies the relative change in the longitudinal coordinate of a particle with a typical energy deviation $\sigma_\delta$ within the bunch over one turn. The rigid-beam approximation assumes that the beam remains frozen when it traverses regions with impedance. For traditional impedances, which typically have wavelengths much larger than $l_s$, lumping such impedances at one point of the entire ring poses no problem for beam dynamics studies. However, for CSR effects, which are typically significant at sub-cm wavelengths, $l_s$ can be comparable to or even larger than CSR wavelengths. In this case, lumping the CSR impedance of the distributed bending magnets into one point is not suitable for CSR instability studies~\cite{OhmiPrivate}.
\end{itemize}
The asymmetric placement of the chamber walls also modifies the CSR impedance models (for example, see~\cite{mori:ipac23-mopa108}), but this aspect is beyond the scope of this paper.

The steady-state 1D model in free space for the CSR impedance with $0<k\ll k_c$ is given by~\cite{faltens1973longitudinal}
\begin{equation}
    Z_\parallel^\text{FS}(k)=
    \frac{(\sqrt{3}+i)Z_0L}{4\cdot 3^{1/3}\pi}
    \Gamma\left( \frac{2}{3} \right)
    \left(\frac{k}{\rho^2}\right)^{1/3},
    \label{eq:csrFS}
\end{equation}
where $L=2\pi\rho$, which assumes that the total bending angle is $2\pi$ with a bending radius of $\rho$ for all dipole magnets. If there are $N$ types of dipole magnets, Eq.~(\ref{eq:csrFS}) remains valid with $L\rho^{-2/3}$ replaced by
\begin{equation}
    \frac{L}{\rho^{2/3}}=
    \sum_{i=1}^N \frac{L_i}{|\rho_i|^{2/3}},
    \label{eq:rhoAverage}
\end{equation}
where the $i$-th type of magnet has a bending radius $\rho_i$ and a total length $L_i$. Here, a negative $\rho_i$ indicates a reverse bend. An average bending radius can be obtained as $L=\rho\theta$ with $\theta=\sum_i L_i/|\rho_i|$.

\begin{figure}
    \centering
    \includegraphics[width=\linewidth]{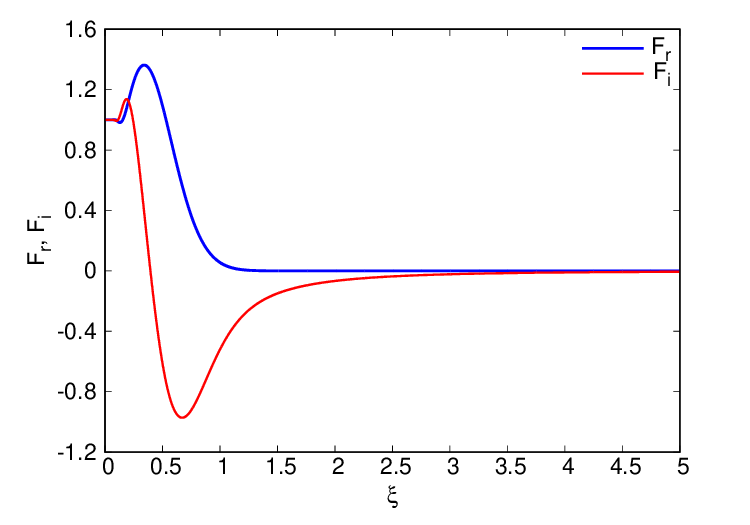}
    \caption{Shielding parameter as a function of $\xi$ for parallel-plates steady-state (PP-SS) CSR impedance.}
    \label{fig:Fpp}
\end{figure}

The steady-state 1D model with parallel-plates shielding and $0<k\ll k_c$ is well approximated in terms of Airy functions by~\cite{agoh2004calculation}
\begin{equation}
    Z_\parallel^\text{PP}(k)=F_\text{r}(\xi)\text{Re}[Z_\parallel^\text{FS}(k)]+
    iF_\text{i}(\xi)\text{Im}[Z_\parallel^\text{FS}(k)],
    \label{eq:csrPP}
\end{equation}
where $\text{Re}[]$ and $\text{Im}[]$ respectively indicate taking the real and imaginary part of a complex number; $F_\text{r}(\xi)$ and $F_\text{i}(\xi)$ represent the shielding parameters due to parallel plates, given by
\begin{equation}
    F_\text{r}(\xi)=
    \frac{g(\xi)}{\sqrt{3}}
    \sum_{p=0}^\infty
    \left[\text{Ai}'^2(\zeta_p^2)+\zeta_p^2 \text{Ai}^2(\zeta_p^2) \right],
    \label{eq:Fr}
\end{equation}
\begin{equation}
    F_\text{i}(\xi)=
    -g(\xi)
    \sum_{p=0}^\infty
    \left[\text{Ai}'(\zeta_p^2)\text{Bi}'(\zeta_p^2)+\zeta_p^2\text{Ai}(\zeta_p^2)\text{Bi}(\zeta_p^2)\right],
    \label{eq:Fi}
\end{equation}
with
\begin{equation}
    g(\xi)=\frac{4\pi^2}{\Gamma\left(\frac{2}{3}\right)} (12\xi^2)^{1/3},
\end{equation}
\begin{equation}
    \xi=\sqrt{\frac{\rho}{2k^2h^3}}=\frac{2k_w}{\pi k},
\end{equation}
\begin{equation}
    \zeta_p=\frac{(2p+1)\pi}{2}\xi^{2/3}.
\end{equation}
With the formulation in Eq.~(\ref{eq:csrPP}), the shielding parameters $F_\text{r,i}(\xi)$ are functions of the dimensionless parameter $\xi$ as shown in Fig.~\ref{fig:Fpp}. In the limit of $\xi\rightarrow 0$ (i.e., when $h\rightarrow \infty$ or large $k$), the summation over $p$ is replaced by an integral and $F_\text{r,i}(\xi)$ approaches 1~\cite{agoh2004dynamics}. In the limit of $\xi\rightarrow \infty$ (i.e., low-frequency region where $k\rightarrow 0$), $F_\text{r,i}(\xi)$ approaches 0 due to strong shielding.

Various theories and codes have been developed for the calculation of CSR impedance in electron storage rings. For theories of steady-state CSR in a rectangular chamber, refer to~\cite{agoh2009steady, cai2014scaling}. For theories of transient CSR with wall shielding, see~\cite{stupakov2016analytical, agoh2023transient}. For theories of steady-state CSR from wigglers, see~\cite{wu2003calculation, blednykh2023microwave}. Numerical codes dedicated to the calculation of the CSR impedance were presented in~\cite{stupakov2009calculation, agoh2004calculation, zhou2011thesis}. In this paper, we mainly use the CSRZ code~\cite{zhou2011thesis} to calculate the CSR impedance. The impedance data are subsequently employed for the analysis and simulation of MWI in electron rings.


\section{\label{sec:instability}Instability analysis}


Instability analysis was performed in~\cite{stupakov2002beam} to determine the MWI threshold arising from high-frequency free-space CSR impedances with $k\sigma_z\gg 1$. Keeping the coasting beam assumption of $k\sigma_z\gg 1$, the theory of~\cite{stupakov2002beam} can be extended to include the arbitrary impedance of $Z_\parallel(k)$~\cite{korostelev2011wake, blednykh2023microwave}. For FS-SS CSR, the explicit formula for the MWI threshold as a function of the CSR wavelength is given by~\cite{blednykh2023microwave}
\begin{equation}
    I_\text{th}(\lambda)=\frac{3^{5/6}(2\pi)^{5/3}}{2\Gamma\left( \frac{2}{3} \right)G_\text{i}}
    \frac{(E/e)\eta\sigma_\delta^2\sigma_z}{Z_0\rho^{1/3}\lambda^{2/3}},
    \label{eq:Ith_CSR_FS}
\end{equation}
where $\lambda=2\pi / k$, and $G_\text{i}$ is a constant (i.e., approximately 1.86 and -1.09 for positive and negative $\eta$, respectively).

Following the formulations in Sec.~IV of~\cite{blednykh2023microwave}, we derive a scaling law for the CSR instability threshold in the case of the PP-SS impedance model. The equation to be solved is
\begin{equation}
    \frac{G_\text{i}(A_\text{th})}{G_\text{r}(A_\text{th})} =\frac{Z_r(k)}{Z_i(k)},
    \label{eq:ThresholdCondition}
\end{equation}
where $Z_r(k)=\text{Re}[Z_\parallel(k)]$, $Z_i(k)=\text{Im}[Z_\parallel(k)]$, and the two functions with real arguments are given by
\begin{equation}
    G_\text{r}(A) =\sqrt{2\pi}-\pi A e^{-A^2/2} \text{erfi}[A/\sqrt{2}],
\end{equation}
\begin{equation}
    G_\text{i}(A) =\text{sgn}[\eta]\pi A e^{-A^2/2},
\end{equation}
where $\text{erfi}[x]$ is the imaginary error function and $\text{sgn}[x]$ denotes the sign function. With impedance $Z_\parallel(k)$ given, Eq.~(\ref{eq:ThresholdCondition}) is solved to find $A_\text{th}$, which is a function of $k$ and determines the threshold current as
\begin{equation}
    f_\text{th}=\frac{kZ_r}{G_\text{i}(A_\text{th})(Z_r^2+Z_i^2)}=\frac{kZ_i}{G_\text{r}(A_\text{th})(Z_r^2+Z_i^2)},
    \label{eq:ThresholdCurrent}
\end{equation}
with
\begin{equation}
    f_\text{th}=\frac{I_\text{th}}{2\pi(E/e)\eta\sigma_\delta^2\sigma_z}.
\end{equation}

Utilizing Eq.~(\ref{eq:csrPP}), the relevant equations are rewritten as
\begin{equation}
    \frac{G_\text{i}(A_\text{th})}{G_\text{r}(A_\text{th})} =\frac{\sqrt{3}F_\text{r}(\xi)}{F_\text{i}(\xi)},
    \label{eq:ThresholdConditionPP}
\end{equation}
\begin{equation}
    f_\text{th}\frac{Z_r}{k}=\frac{3F_\text{r}}{G_\text{i}(A_\text{th})(3F_\text{r}^2+F_\text{i}^2)}.
    \label{eq:ThresholdCurrentPP1}
\end{equation}
Substituting Eq.~(\ref{eq:csrFS}) into Eq.~(\ref{eq:ThresholdCurrentPP1}), we obtain
\begin{equation}
    I_\text{th}=
    \frac{4\pi(E/e)\eta\sigma_\delta^2\sigma_z^{1/3}}{Z_0\rho^{1/3}}
    \frac{\Pi^{2/3}}{12^{1/6}\Gamma\left(\frac{2}{3}\right)}
    F_\text{th}(\xi),
    \label{eq:ThresholdCurrentPP2}
\end{equation}
with
\begin{equation}
    F_\text{th}(\xi)=
    \frac{3F_\text{r}}{G_\text{i}(A_\text{th})(3F_\text{r}^2+F_\text{i}^2)\xi^{2/3}},
\end{equation}
where $k=\Pi/(\sqrt{2}\xi\sigma_z)$ is utilized. For FS-SS CSR, Eq.~(\ref{eq:ThresholdCurrentPP2}) reduces to Eq.~(\ref{eq:Ith_CSR_FS}) with $F_\text{th}=3/(4G_\text{i}\xi^{2/3})$. Equation~(\ref{eq:ThresholdCurrentPP2}) is an extension of Eq.~(\ref{eq:Ith_CSR_FS}) from free space to the parallel-plates model of steady-state CSR. The threshold current is inherently dependent on the CSR wavelength, as revealed by the nature of instability analysis employing impedance.
\begin{figure}
    \centering
    \includegraphics[width=\linewidth]{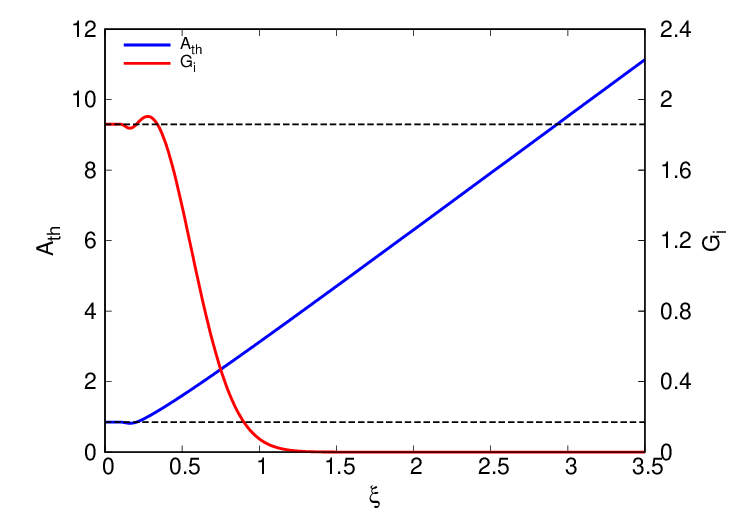}
    \caption{$A_\text{th}(\xi)$ and $G_\text{i}(A_\text{th})$ for PP-SS CSR. The dashed black lines indicate $A_\text{th}\approx 0.85$ and $G_\text{i} \approx 1.86$, respectively, corresponding to FS-SS CSR~\cite{blednykh2023microwave}.}
    \label{fig:AthPP}
\end{figure}

\begin{figure}
    \centering
    \includegraphics[width=\linewidth]{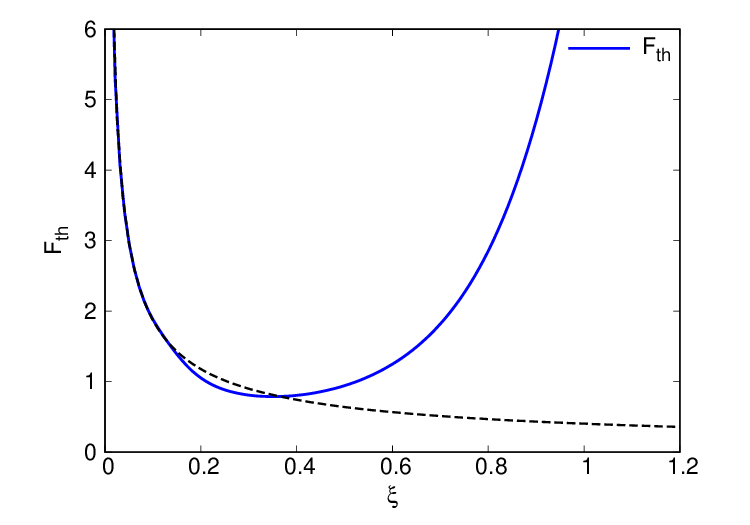}
    \caption{$F_\text{th}(\xi)$ for PP-SS CSR. The dashed black line indicates $F_\text{th}(\xi)=3/(4G_\text{i}\xi^{2/3})$ with $G_\text{i}\approx 1.86$ for the FS-SS CSR~\cite{blednykh2023microwave}.}
    \label{fig:FthPP}
\end{figure}

Given $F_\text{r,i}$ by Eqs.~(\ref{eq:Fr}) and~(\ref{eq:Fi}) (also see Fig.~\ref{fig:Fpp}), we are to solve Eq.~(\ref{eq:ThresholdConditionPP}) numerically and find $A_\text{th}$ (and $G_\text{i}$) as a function of $\xi$, as shown in Fig.~\ref{fig:AthPP}. Consequently, $F_\text{th}(\xi)$ is calculated as shown in Fig.~\ref{fig:FthPP}, which clearly shows a minimum value $\text{Min}[F_\text{th}(\xi)]$ at $\xi_\text{th}$. This value gives the minimum bunch current at which the beam will become unstable. Similar to Eq.~(\ref{eq:IthCSR}), we achieve an alternative scaling law of
\begin{equation}
    I_\text{th2}=
    \frac{4\pi(E/e)\eta\sigma_\delta^2\sigma_z^{1/3}}{Z_0\rho^{1/3}}
    S_\text{th2},
    \label{eq:IthCSR2}
\end{equation}
with
\begin{equation}
    S_\text{th2}=
    \frac{\Pi^{2/3}}{12^{1/6}\Gamma\left(\frac{2}{3}\right)}
    \text{Min}[F_\text{th}(\xi)].
\end{equation}
By numerical calculations, we find $\xi_\text{th}\approx 0.35$ and $\text{Min}[F_\text{th}(\xi)]\approx 0.79$, leading to
\begin{equation}
    S_\text{th2}\approx
    0.384\Pi^{2/3}.
    \label{eq:Sth2}
\end{equation}

\begin{figure}
    \centering
    \includegraphics[width=\linewidth]{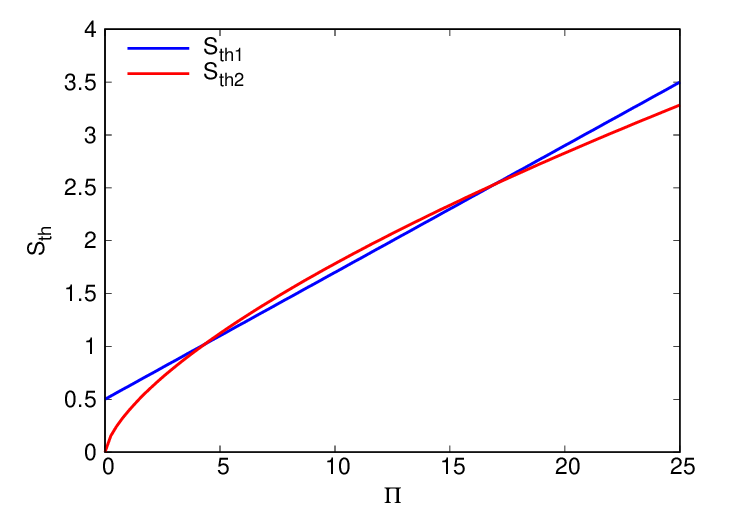}
    \caption{Comparison of $S_\text{th1}(\Pi)$ by Eq.~(\ref{eq:Sth1}) and $S_\text{th2}(\Pi)$ by Eq.~(\ref{eq:Sth2}) for PP-SS CSR.}
    \label{fig:SthComparison}
\end{figure}

Equation~(\ref{eq:Sth2}) holds true under the assumptions of $k\sigma_z\gg 1$, which is a prerequisite for the entire instability analysis. $\xi_\text{th}\approx 0.35$ determines a critical wavenumber of
\begin{equation}
    k_\text{th}\approx 2.0\sqrt{\rho/h^3}=2.0\Pi/\sigma_z.
    \label{eq:kth}
\end{equation}
The CSR field at this wavenumber is the most potent in driving instability and, consequently, is the most detrimental. This suggests that the validity condition of Eq.~(\ref{eq:Sth2}) is $\Pi\gg 0.5$. In the end, we compare Eqs.~(\ref{eq:Sth1}) and~(\ref{eq:Sth2}) in Fig.~\ref{fig:SthComparison}. For $\Pi\gg 0.5$, over a large range, the two theories, one derived from simulations and the other from theoretical analysis, exhibit small discrepancies.

Cai performed a similar analysis for the PP-SS CSR in~\cite{cai:ipac11-frxaa01}, and our findings are consistent with that work. For example, Eqs.~(\ref{eq:IthCSR2}) and~(\ref{eq:kth}) correspond to Eqs.~(12) and~(11) of~\cite{cai:ipac11-frxaa01}, respectively. For Eq.~(\ref{eq:Sth2}), Cai found $S_\text{th}=3\Pi^{2/3}/(\sqrt{2}\pi^{3/2})\approx 0.381\Pi^{2/3}$, which is very close to Eq.~(\ref{eq:Sth2}). He also noted that the CSR instability threshold follows $I_\text{th2}\propto \gamma\eta\sigma_\delta^2\sigma_z/h$ and is independent of $\rho$ when $\Pi\gg 0.5$. We emphasize that we are solving Eq.~(\ref{eq:ThresholdCondition}) to determine the instability threshold, rather than solving the original dispersion-relation equation. The primary difference lies in solving a single-variable equation on the real axis rather than a complex-variable equation involving special functions.

\begin{figure}
    \centering
    \includegraphics[width=\linewidth]{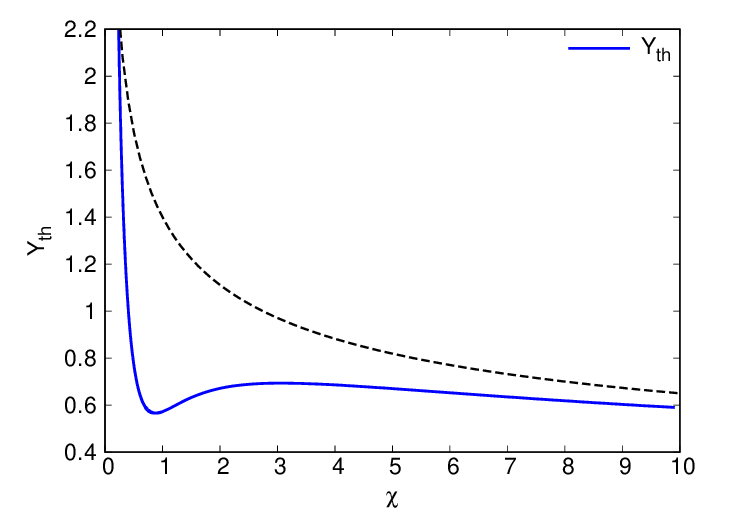}
    \caption{$Y_\text{th}(\chi)$ for resistive wall impedance of Eq.~(\ref{eq:Zrw}). The dashed black line indicates $Y_\text{th}=1/(0.714\chi^{1/3})$ in the limit of $\chi\rightarrow \infty$.}
    \label{fig:YthRW}
\end{figure}

The scaling laws discussed in this section offer a quick check of the CSR instability for electron storage rings. In practice, it is advisable to directly solve Eqs.~(\ref{eq:ThresholdCondition}) and~(\ref{eq:ThresholdCurrent}) using an arbitrary $Z_\parallel(k)$ obtained from bottom-up impedance modelings (for examples of this approach, see~\cite{blednykh2021impedance, wang2022broadband}):
\begin{itemize}
    \item The CSR impedance model may deviate from the PP-SS model due to practical configurations in optics design and vacuum chambers.
    \item Wigglers and undulators can also contribute to CSR impedance, especially when strong wigglers are used heavily for beam-emittance damping~\cite{wu2003calculation, blednykh2023microwave}.
    \item There may be other vacuum components that contribute to the high-frequency impedance. In particular, the resistive wall (RW) impedance may be important in modern low-emittance electron storage rings (for relevant examples, see~\cite{migliorati2018impact, li2024terahertz}), where reducing chamber apertures follows the trend of squeezing the beam emittance for higher brightness.
\end{itemize}
Assume that a storage ring has a smooth chamber with length $L_\text{RW}$ and full height of $2h$, then the resistive wall impedance is given by~\cite{chao1993physics}
\begin{equation}
    Z_\parallel^\text{RW}(k)=
    \frac{f_YZ_0L_\text{RW}}{\pi h\left( 2\sqrt{\frac{iZ_0\sigma_c}{k}} -ihk \right)},
    \label{eq:Zrw}
\end{equation}
where $\sigma_c$ is the electrical conductivity of the chamber material, and $f_Y$ is the Yokoya factor: 1 for a round chamber and $\pi^2/12$ for parallel plates. Substituting Eq.~(\ref{eq:Zrw}) into Eq.~(\ref{eq:ThresholdCondition}), we obtain
\begin{equation}
    \frac{G_\text{i}(A_\text{th})}{G_\text{r}(A_\text{th})} =
    \frac{\chi}{1-\chi},
    \label{eq:ThresholdConditionRW1}
\end{equation}
with the dimensionless scaling factor
\begin{equation}
    \chi=
    \frac{\sqrt{2Z_0\sigma_c}}{hk^{3/2}}.
\end{equation}
Applying the solution of Eq.~(\ref{eq:ThresholdConditionRW1}) to Eq.~(\ref{eq:ThresholdCurrent}), we derive the resistive-wall instability threshold as
\begin{equation}
    I_\text{th3}=
    \frac{2\pi^2(E/e)\eta\sigma_\delta^2\sigma_z(2Z_0\sigma_ch)^{2/3}}{f_YZ_0L_\text{RW}}
    \text{Min}[Y_\text{th}(\chi)],
    \label{eq:ThresholdCurrentRW3}
\end{equation}
with
\begin{equation}
    Y_\text{th}(\chi)=
    \frac{1}{G_\text{i}(A_\text{th})\chi^{1/3}}.
\end{equation}
In the limit of $\chi\rightarrow \infty$, the solution of Eq.~(\ref{eq:ThresholdConditionRW1}) is $A_\text{th}\approx 2.12$, leading to $G_\text{i}(A_\text{th})\approx 0.714$. The function $Y_\text{th}(\chi)$, as plotted in Fig.~\ref{fig:YthRW}, has a minimum value of $\text{Min}[Y_\text{th}(\chi)]\approx 0.566$ at $\chi_\text{th}\approx 0.877$, which determines a critical wavenumber of
\begin{equation}
    k_\text{th}^\text{RW}=
    \left( \frac{2Z_0\sigma_c}{\chi_\text{th}^2h^2}\right)^{1/3}
    \approx 1.37 \left( \frac{Z_0\sigma_c}{h^2}\right)^{1/3}.
\end{equation}
The validity condition of Eq.~(\ref{eq:ThresholdCurrentRW3}) with $\text{Min}[Y_\text{th}(\chi)]\approx 0.566$ is $k_\text{th}^\text{RW}\sigma_z\gg 1$, corresponding to
\begin{equation}
    \Pi_\text{RW}=\sigma_z \left( \frac{Z_0\sigma_c}{h^2}\right)^{1/3} \gg 0.73.
\end{equation}

Microwave instability driven by high-frequency RW impedance was recently reported in~\cite{li2024terahertz}, where the RW impedance with NEG coating was utilized in the tracking simulations to detect the instability. The instability analysis developed in this section can be applied to validate such simulations.

Table~\ref{tab:parameters} summarizes the machine parameters of two different low-emittance electron rings: Elettra 2.0, representative of ring-based light sources, and the SuperKEKB LER, representative of e+e- colliders. The derived parameters relevant to the CSR and RW instability are also listed; however, we postpone the discussion of physics to the following two sections, where more detailed calculations and analyses are presented. It should be noted that for $I_{th3}$ in Table~\ref{tab:parameters}, we assume a simplified case for Elettra 2.0 where the entire vacuum chamber is made of stainless steel with an electrical conductivity of $\sigma_c=1.45\times 10^6$ S/m. To estimate $I_{th3}$ for SuperKEKB LER, it is assumed that all vacuum chambers are made of aluminum with $\sigma_c=3.8\times 10^7$ S/m. In practical RW impedance modeling, we consider realistic configurations of chambers with coatings for these projects. See the following sections for further details.
\begin{table}[]
\caption{Machine parameters for Elettra 2.0 with and without a third-order harmonic cavity (3HC) and SuperKEKB LER (SLER), with calculated parameters refer to the definitions in this paper.}
\label{tab:parameters}
\resizebox{\columnwidth}{!}{%
\begin{tabular}{|c|cc|c|c|}
\hline
\textbf{Paremeter}                           & \multicolumn{2}{c|}{\textbf{Elettra 2.0}}   & \textbf{SLER}     & \textbf{Unit}         \\ \hline
                                             & \multicolumn{1}{c|}{Without 3HC} & With 3HC &                       & -                     \\ \hline
Energy ($E$)                                   & \multicolumn{2}{c|}{2.4}                 &         {4.0}              & GeV                    \\ \hline
Circumference ($C$)                            & \multicolumn{2}{c|}{259.2}                  &        {3016.3}               & m                     \\ \hline
\begin{tabular}[c]{@{}c@{}}Bunch length\\ ($\sigma_{z}$)\end{tabular} &
  \multicolumn{1}{c|}{1.8} &
  4.5 &    {4.6}
   &
  mm \\ \hline
\begin{tabular}[c]{@{}c@{}}Synchrotron Tune\\ ($Q_{s}$)\end{tabular} &
  \multicolumn{1}{c|}{2.6e-3} &  
   &  {0.023}
   &
  - \\ \hline
Energy spread ($\sigma_{p}$)     & \multicolumn{2}{c|}{9.1e-4}               &       {7.53e-4}                & -                     \\ \hline
Longitudinal damping time ($\tau_{\delta}$)     & \multicolumn{2}{c|}{6.7}               &        {22.9}               & ms                    \\ \hline
Compaction factor ($\alpha_{c}$) & \multicolumn{2}{c|}{1.2e-4}                 &        {2.97e-4}               & -                     \\ \hline
Bending radius ($\rho$) &
  \multicolumn{2}{c|}{\begin{tabular}[c]{@{}c@{}}Type1: 10.19\\ Type 2: 7.05\\ Type 3(anti-bend): 34.37\\ average(type1\&2)=7.81\end{tabular}} &     {74.7}
   &
  m \\ \hline
Bending length (L) &
  \multicolumn{2}{c|}{\begin{tabular}[c]{@{}c@{}}Type 1: 0.64\\ Type 2: 0.80\\ Type 3 (anti-bend): 0.24\end{tabular}} &
  {4.2} &    
  m \\ \hline
Vacuum chamber size ($h$)                      & \multicolumn{2}{c|}{7.5}                 &         {45}              & mm                     \\ \hline
Critical wavenumber ($k_{c}$)                        & \multicolumn{2}{c|}{2.0e+10}             & 9.6e+9 & $\text{m}^{-1}$ \\ \hline
Wall shielding threshold ($k_{w}$)                    & \multicolumn{2}{c|}{4.8e+3}              & \multicolumn{1}{l|}{1.0e+3} & $\text{m}^{-1}$ \\ \hline
Radiation formation length ($l_{f}$)                & \multicolumn{1}{c|}{1.4}      & 1.9   & \multicolumn{1}{c|}{8.5} & m                     \\ \hline
Catch-up distance ($l_{c}$)                            & \multicolumn{2}{c|}{0.42}                 & \multicolumn{1}{c|}{5.2} & m \\ \hline
Slippage length ($l_{s}$)                              & \multicolumn{2}{c|}{2.8e-5}                       & \multicolumn{1}{c|}{6.7e-4} & m \\ \hline
\begin{tabular}[c]{@{}c@{}} $\Pi$\end{tabular} &
  \multicolumn{1}{c|}{7.5} &
  19.4 &    {4.2}
   &
  - \\ \hline
\begin{tabular}[c]{@{}c@{}} $\Pi_\text{RW}$\end{tabular} &
  \multicolumn{1}{c|}{34.4} &
  96.0 &  {88.3}
   &
  - \\ \hline
\begin{tabular}[c]{@{}c@{}} $I_\text{th1}$ \end{tabular} &
  \multicolumn{1}{c|}{0.7} &
  1.9 &  0.89
   &
  mA \\ \hline
\begin{tabular}[c]{@{}c@{}} $I_\text{th2}$ \end{tabular} &
  \multicolumn{1}{c|}{0.73} &
  1.8 &  0.88
   &
  mA \\ \hline
\begin{tabular}[c]{@{}c@{}} $I_\text{th3}$ \end{tabular} &
  \multicolumn{1}{c|}{2.4} &
  6.1 &  36
   &
  mA \\ \hline
\end{tabular}%
}
\end{table}

\section{\label{sec:csrElettra2}CSR effects in Elettra 2.0}

\subsection{\label{sec:impElettra2} Longitudinal broadband impedance model}

As is common for low-emittance ring-based light sources, the primary sources of impedance in Elettra 2.0~\cite{karantzoulis2024design} include resistive walls, transitions and tapers, bellows, BPMs, strip lines, RF cavities, flanges, antechambers, and kickers. The broadband impedance model for Elettra 2.0 is reviewed in~\cite{karantzoulis2021elettra}. Here, we summarize the main results relevant to our study of the CSR instability.

The typical vacuum chamber of Elettra 2.0 is rhomboidal with internal dimensions 27$\times$17 mm. In practice, the IW2D code~\cite{mounet2010electromagnetic, mounet2012lhc} is used to calculate the RW impedance with a rectangular approximation of the vacuum chamber, considering the NEG coating. The chamber materials include copper (45\%), aluminum (20\%), and stainless steel (35\%). For the pessimistic approximation in IW2D, the NEG coating of 0.5 $\mu$m is considered.

The broadband impedances from geometric discontinuities in various vacuum components are estimated using analytic formulas. The results are used to construct a broadband resonator (BBR) model as
\begin{equation}
    Z_{||}(k) =\frac{R_{s}}{1+iQ(\frac{k_{r}}{k}-\frac{k}{k_{r}})},
    \label{eq:bbr}
\end{equation}
with quality factor $Q=1$ and the resonant frequency $f_r=k_rc/(2\pi)=7$ GHz. Here, $f_r$ is chosen to be the cutoff frequency of the regular vacuum chamber, and $R_s$ is estimated from $|\text{Im}[Z_\parallel]/n|=0.5\ \Omega$ according to Table 4.1.5 of~\cite{karantzoulis2021elettra}. According to Eq.~(\ref{eq:bbr}), this corresponds to $\omega_0R_s/(Qk_rc)=0.5\ \Omega$, resulting in $R_s=3026.1$ $\Omega$. The $Z_\parallel/n$ of the RW impedance is excluded, as we will use the IW2D data directly.

The CSRZ code~\cite{zhou2011thesis} is used to calculate the CSR impedance of three types of bending magnets:
\begin{itemize}
    \item 24 normal-type bends with 0.64 m length and $3.6^\circ$ per magnet.
    \item 48 longitudinal-gradient bends (LGBs) with 0.8 m length and $6.5^\circ$ per magnet.
    \item 96 combined-function magnets, which have dipole and quadrupole fields, with 0.24 m length and $-0.4^\circ$ per magnet.
\end{itemize}
For a typical calculation, a single bend connected with an infinitely long drift is used for the CSRZ model. A rectangular chamber with full height $2h$ = 15 mm and full width $2w$ = 27 mm is used to approximate the boundary of the metal walls. Multi-bend interference~\cite{zhou2011calculation} between the CSR fields of consecutive bends is neglected.
\begin{figure}
\centering
\includegraphics[width=\linewidth]{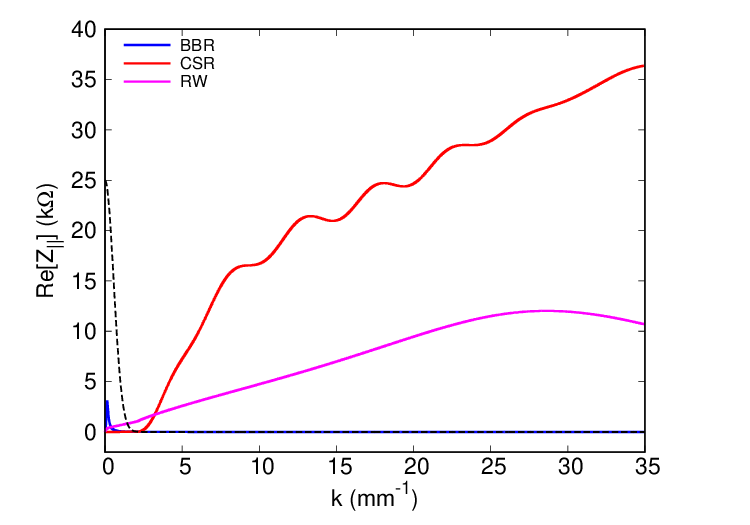}
\includegraphics[width=\linewidth]{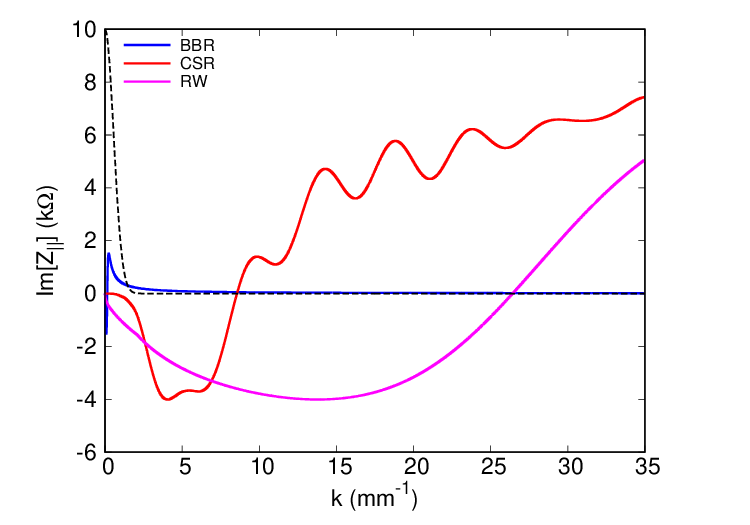}
\caption{Comparison of real and imaginary parts of BBR, CSR and RW impedances for Elettra 2.0. The black dashed lines show the beam spectrum of the nominal Gaussian bunch with $\sigma_{z}=1.8$ mm.}
\label{fig:ZL-Elettra2.0}
\end{figure}

The three impedances mentioned above for Elettra 2.0 are contrasted in Fig.~\ref{fig:ZL-Elettra2.0}. The maximum wavenumber for the impedance calculations is set to $k_\text{max}=35000\ \text{m}^{-1}$, which corresponds to $k_\text{max}\sigma_z=63$ with $\sigma_z=1.8$ mm (the bunch length at the zero-current limit). This choice should be sufficient for the analysis and simulations of MWI. It is evident that the CSR and RW impedances dominate the high-frequency region where $k\gg 1/\sigma_z$. Inductive and resistive impedances at $ k \lesssim 1/\sigma_z $ contribute to bunch lengthening and profile tilting by distorting the potential well. These effects may influence the determination of the CSR instability threshold, as will be discussed later.


\subsection{\label{sec:anaElettra2}Instability analysis}

\begin{figure}
\centering
\includegraphics[width=\linewidth]{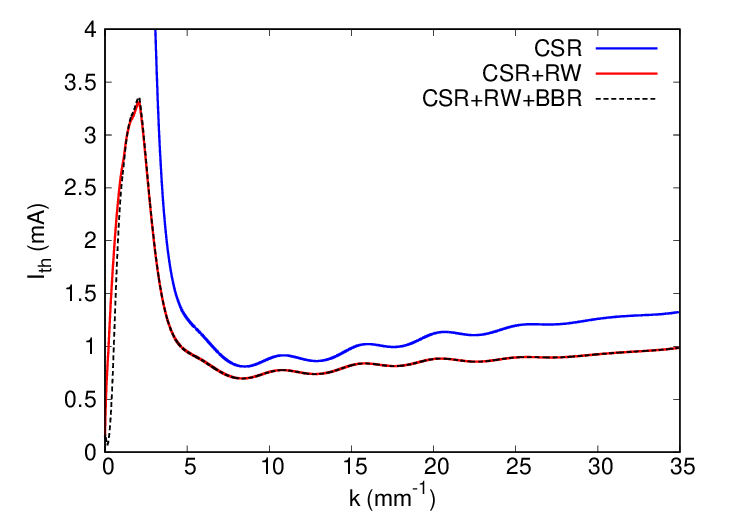}
\caption{Threshold current as a function of wavenumber for Elettra 2.0 with nominal bunch length $\sigma_z$=1.8 mm. The impedance models for the analysis are provided in Fig.~\ref{fig:ZL-Elettra2.0}. Additional beam parameters are given in Tab.~\ref{tab:parameters}.}
\label{fig:Ibth-Elettra2.0}
\end{figure}

The influence of high-frequency CSR and RW impedances on MWI can be explored through the instability analysis described in Section~\ref{sec:instability}. Using the Elettra 2.0 parameter set with $\sigma_z$=1.8 mm in Tab.~\ref{tab:parameters}, we perform an analysis using the impedance data of Fig.~\ref{fig:ZL-Elettra2.0} and obtain the threshold bunch current as a function of the wavenumber as shown in Fig.~\ref{fig:Ibth-Elettra2.0}. As required by the validity condition of the theory, we search for the minimum threshold current in the region of $k\gg 1/\sigma_z\approx$ 560 $\text{m}^{-1}$: $I_\text{th}\approx$ 0.81 mA at $k_\text{th}\approx 8500$ $\text{m}^{-1}$ when considering only CSR, and $I_\text{th}\approx$ 0.70 mA at $k_\text{th}\approx 8400$ $\text{m}^{-1}$ when considering both CSR and RW. The CSR instability threshold estimated using impedance data obtained by CSRZ is close to that estimated using the PP-SS model (see Table~\ref{tab:parameters}), suggesting that the PP-SS model is quite suitable for CSR instability studies for Elettra 2.0. Figure~\ref{fig:Ibth-Elettra2.0} illustrates that the RW impedance remarkably lowers the MWI threshold, which is also in line with the simple estimate provided in Table~\ref{tab:parameters}. The BBR impedance does not affect the threshold current at $k\gg 1/\sigma_z$. This is expected since the BBR does not contain significant high-frequency impedances.

Figure~\ref{fig:Ibth-Elettra2.0} indicates that, for an accurate detection of the MWI threshold with zero-current limit bunch length $\sigma_z=1.8$ mm, it is essential to appropriately model high-frequency impedances in the region of $9 \lesssim k\sigma_z \lesssim 27$ in numerical simulations. In particular, for tracking simulations using macroparticles, the mesh size used to sample the longitudinal profile of the bunch should satisfy $\Delta z\ll 2\pi/k_\text{max}\approx 0.18$ mm when fully utilizing the impedance data from Fig.~\ref{fig:ZL-Elettra2.0}. Consequently, the number of macroparticles should be sufficiently large to minimize statistical errors in the bunch histogram, which could otherwise magnify impedance effects at high frequencies and induce numerical instability.
\begin{figure}
\centering
\includegraphics[width=\linewidth]{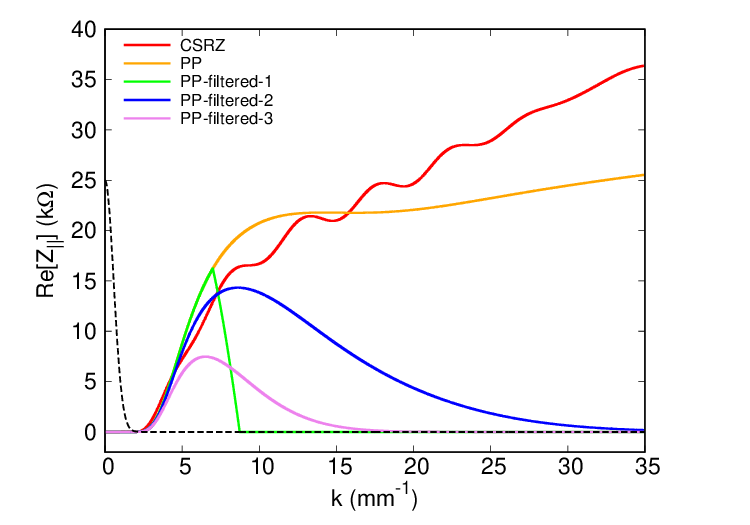}
\includegraphics[width=\linewidth]{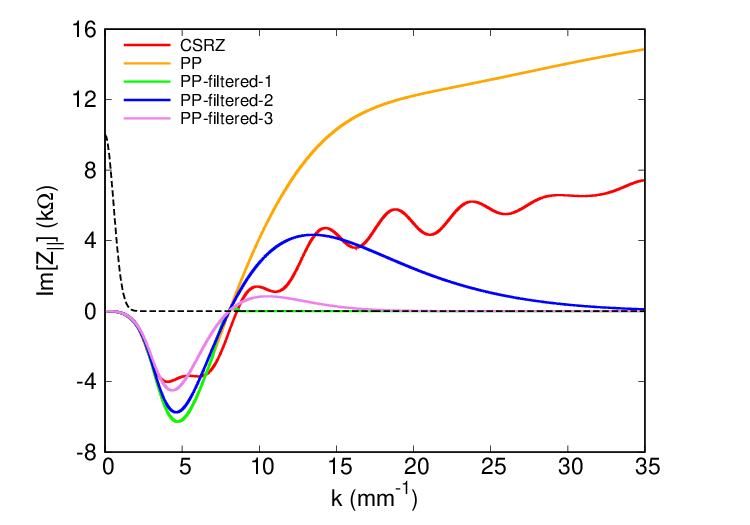}
\caption{Alternative CSR impedance models for Elettra 2.0. Red lines: Computed by CSRZ code; Orange lines: PP-SS model calculated using the \textit{csrImpedance} module of the \textit{Elegant} code; Green lines: PP-SS model with filtering using the option of \textit{-filter=0.2,0.25} of \textit{csrImpedance} module; Blue lines: PP-SS model with Gaussian filter $\text{Exp}[-k^2\hat{\sigma}_{1}^2/2]$; Violet lines: PP-SS model with Gaussian filter $\text{Exp}[-k^2\hat{\sigma}_{2}^2/2]$. The black dashed lines show the beam spectrum of the nominal Gaussian bunch with $\sigma_{z}=1.8$ mm. $(\hat{\sigma}_{1}, \hat{\sigma}_{2})=(0.09, 0.18)$ mm.}
\label{fig:CSR-models-Elettra2.0}
\end{figure}

\begin{figure}
\centering
\includegraphics[width=\linewidth]{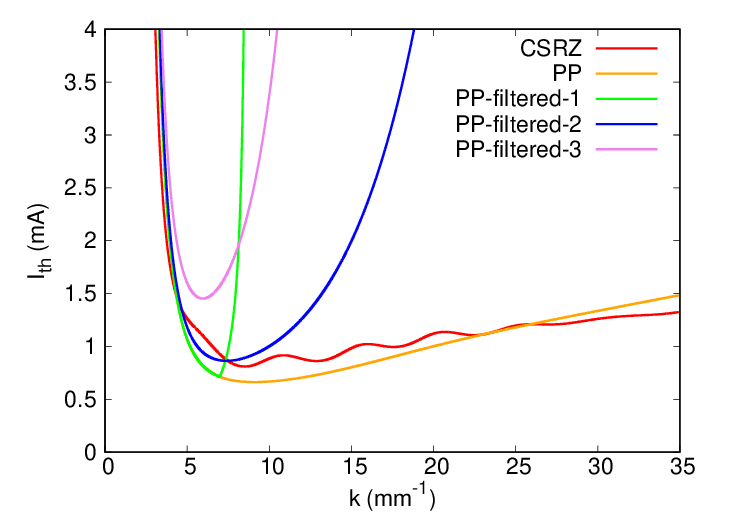}
\caption{Threshold current of CSR instability as a function of wavenumber for Elettra 2.0 with nominal bunch length $\sigma_z$=1.8 mm. The impedance models for the analysis are provided in Fig.~\ref{fig:CSR-models-Elettra2.0}. Additional beam parameters are given in Tab.~\ref{tab:parameters}.}
\label{fig:Ibth-CSR-Elettra2.0}
\end{figure}

Figure~\ref{fig:Ibth-Elettra2.0} also suggests that the high-frequency impedances at $k\sigma_z > 27$ are not crucial for determining the MWI threshold. To save computing resources resulting from small mesh sizes and large numbers of macroparticles in MWI simulations, these high-frequency impedances can be fairly damped by appropriate tapering or filtering. In such instances, instability analysis is useful in establishing criteria to justify the selected filtering functions. Figure~\ref{fig:CSR-models-Elettra2.0} shows alternative CSR impedance models for Elettra 2.0. The corresponding CSR instability threshold current as a function of the wavenumber is shown in Fig.~\ref{fig:Ibth-CSR-Elettra2.0}. The PP-SS model predicts a minimum threshold current of $I_\text{th}\approx 0.66$ mA at $k_\text{th}\approx$ 9100 $\text{m}^{-1}$, slightly lower than the prediction of the CSRZ model. It is important to note that filtering impedance data significantly affects MWI threshold predictions, which will be explored further in the upcoming MWI simulations.

\subsection{\label{sec:simElettra2}Simulations of microwave instability}

There are two simulation methods for microwave instability: the VFP solver (for example, see~\cite{venturini2002bursts}) and the macroparticle tracking. A comprehensive comparison of these methods is provided in~\cite{migliorati2007simulations}. In general, when simulating MWI driven by high-requency impedances, such as CSR and RW impedances, in the region of $k\sigma_z\gg 1$, it is crucial to carefully manage numerical noises to prevent the occurrence of numerical instabilities that could lead to unphysical results.

For Elettra 2.0, we use mainly the Parallel Elegant code~\cite{borland2000elegant} in the tracking simulations. A convergence study was first performed to optimize the simulation setup. We found that for reliable results, macroparticles of $10^{6}$, bin size $2\times 10^{-13}$ s, and number of bins 1024 are necessary when considering the impedance spectrum up to $k_\text{max}=35\ \text{mm}^{-1}$. Due to the short initial bunch length, the case without 3HC in Tab.~\ref{tab:parameters} is the most challenging to simulate. For this case, we also use the VFP solver described in~\cite{bane2010threshold} as a benchmark.

\subsubsection{Comparison of alternative CSR impedance models}

\begin{figure}[hbt!]
    \centering
    \includegraphics[width=\linewidth]{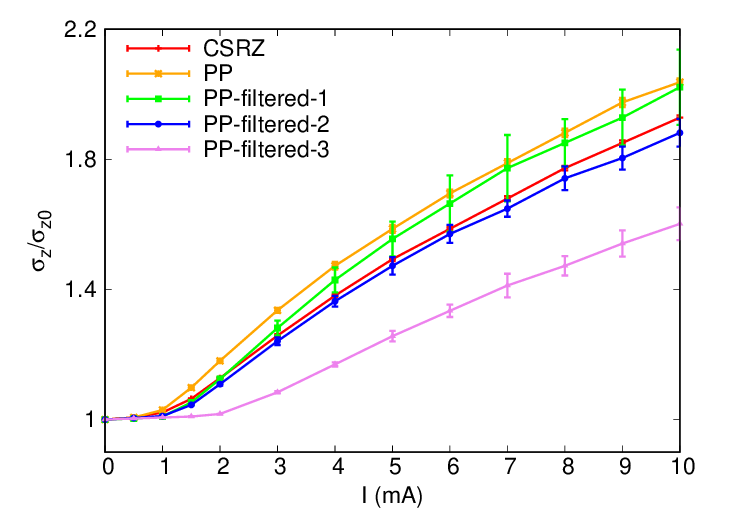}
    \includegraphics[width=\linewidth]{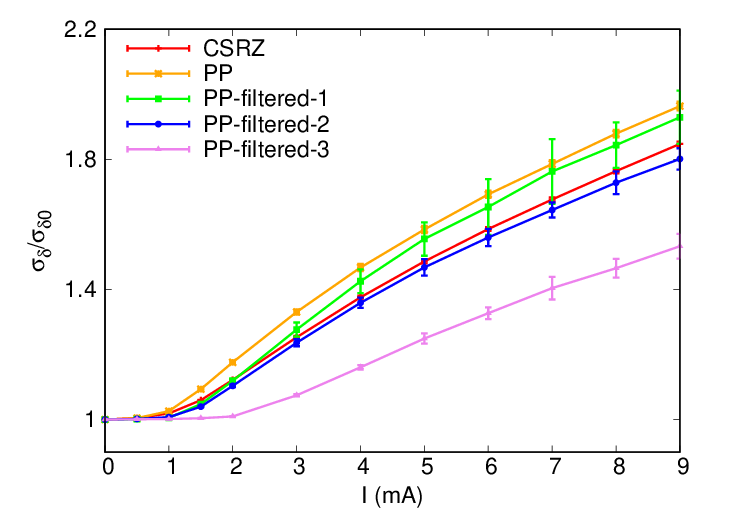}
    \caption{Normalized bunch length and energy spread as a function of bunch current for Elettra 2.0. Simulations were done using Elegant with the impedance models of Fig.~\ref{fig:CSR-models-Elettra2.0}. The bunch is with $10^6$ multi-particles and tracking is done for five times of longitudinal damping time. This result is without 3HC effect.}
    \label{fig:BL_ES_CSR_Elegant}
\end{figure}

\begin{figure}[hbt!]
\centering
\includegraphics[width=\linewidth]{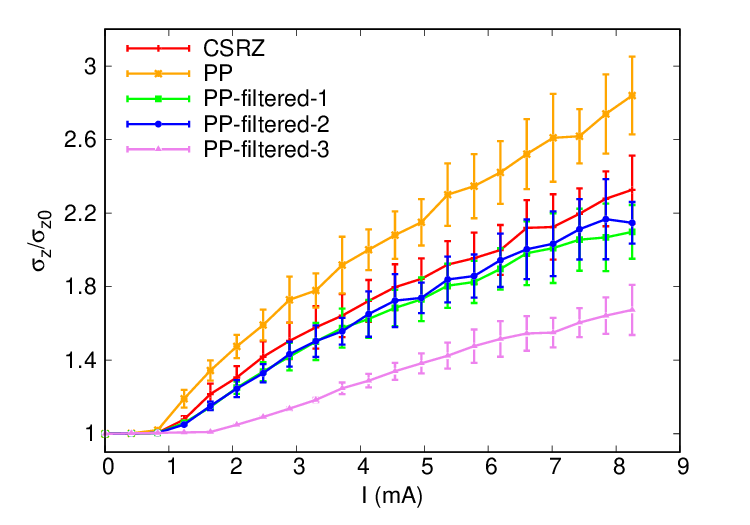}
\includegraphics[width=\linewidth]{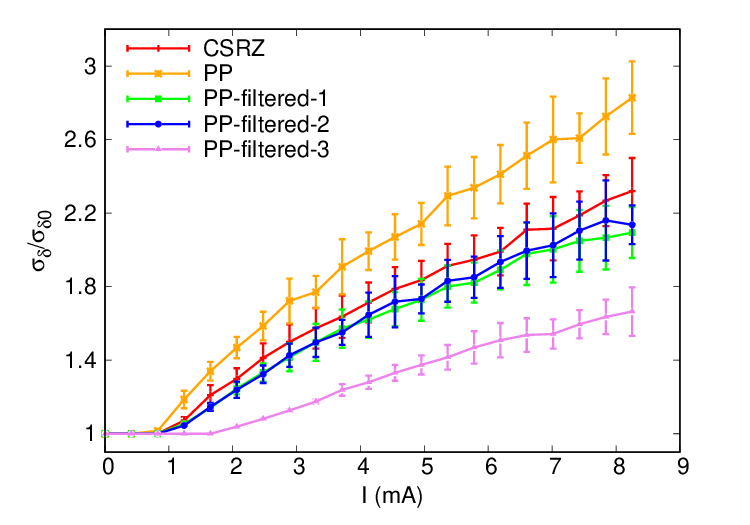}
\caption{Normalized bunch length and energy spread as a function of bunch current for Elettra 2.0. Simulations were done using the VFP solver described in~\cite{bane2010threshold} with the impedance models of Fig.~\ref{fig:CSR-models-Elettra2.0}. Initial beam parameters refer to Tab.~\ref{tab:parameters} without 3HC.}
\label{fig:VFP-CSR-Elettra2.0}
\end{figure}

First, we compare the simulation results using the CSR impedance models in Fig.~\ref{fig:CSR-models-Elettra2.0}. The results without 3HC using the Elegant code and the VFP solver are presented in Figs.~\ref{fig:BL_ES_CSR_Elegant} and~\ref{fig:VFP-CSR-Elettra2.0}, respectively.
The two codes exhibit nearly identical MWI thresholds for the same impedance models. However, the increases in energy spread and bunch length above the MWI threshold differ remarkably. This variation may be attributed to the different smoothing schemes used in the processing of the beam histograms. Moreover, the simulated MWI thresholds are always higher than the theoretical estimates obtained in Fig.~\ref{fig:Ibth-CSR-Elettra2.0}. This can be explained by the inclusion of radiation damping in simulations, a factor that is not considered in instability analyses.

\begin{figure}[hbt!]
    \centering
    \includegraphics[width=\linewidth]{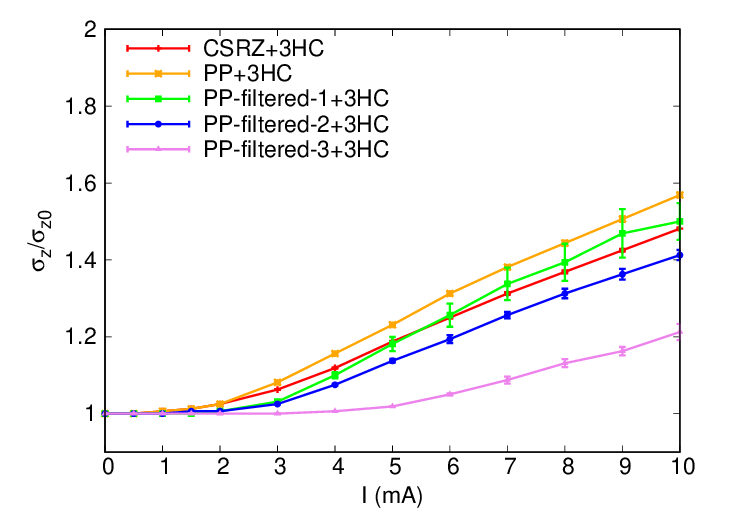}
    \includegraphics[width=\linewidth]{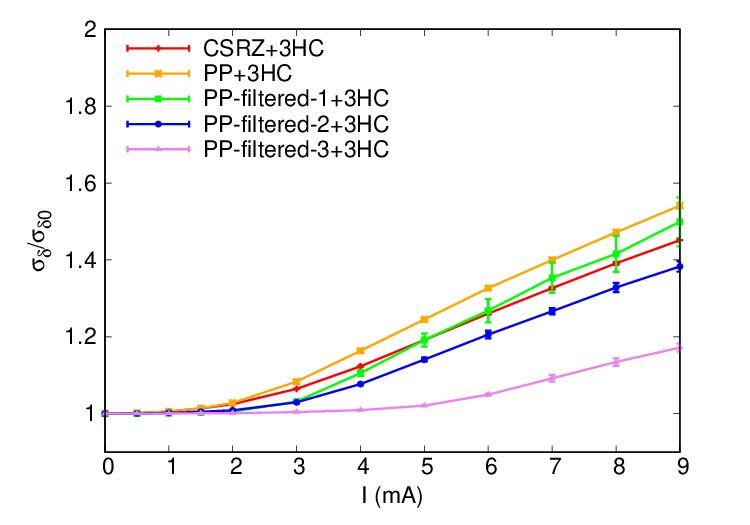}
    \caption{Normalized bunch length and energy spread as a function of bunch current for Elettra 2.0. Simulations were done using Elegant with the impedance models of Fig.~\ref{fig:CSR-models-Elettra2.0}. The bunch is with $10^6$ multi-particles and tracking is done for five times of longitudinal damping time. This result is with 3HC effect.}
    \label{fig:BL_ES_CSR_3HC-Elegant}
\end{figure}

In subsequent simulations, the 3HC is taken into account, and the results of Elegant are depicted in Fig.~\ref{fig:BL_ES_CSR_3HC-Elegant}. Consistent with theoretical expectations, the inclusion of 3HC leads to a reduction in peak current, consequently elevating the MWI threshold by approximately a factor of 2 across all considered CSR impedance models.

Despite the variance in the arbitrary values, both the instability analysis and simulations affirm the necessity for proper treatment of high-frequency CSR impedances when predicting the MWI instability threshold. The impedance model with a Gaussian filter of $\text{Exp}[-k^2\hat{\sigma}^2/2]$, as illustrated in Fig.~\ref{fig:CSR-models-Elettra2.0}, is equivalent to a pseudo-Green function wake with a short bunch having an rms size of $\hat{\sigma}$. Our findings indicate that, to prevent inaccuracies in the predictions, $\hat{\sigma}\lesssim \sigma_z/20$ may be required to construct the pseudo-Green function wake when high-frequency impedances, such as CSR and RW, are incorporated into MWI simulations. This assertion is further supported by the investigations of~\cite{li2024terahertz}.

\subsubsection{\label{sec:CSR+RW+BBRElettra2}Interplay of CSR, RW and BBR impedances}

\begin{figure}
    \centering
    \includegraphics[width=\linewidth]{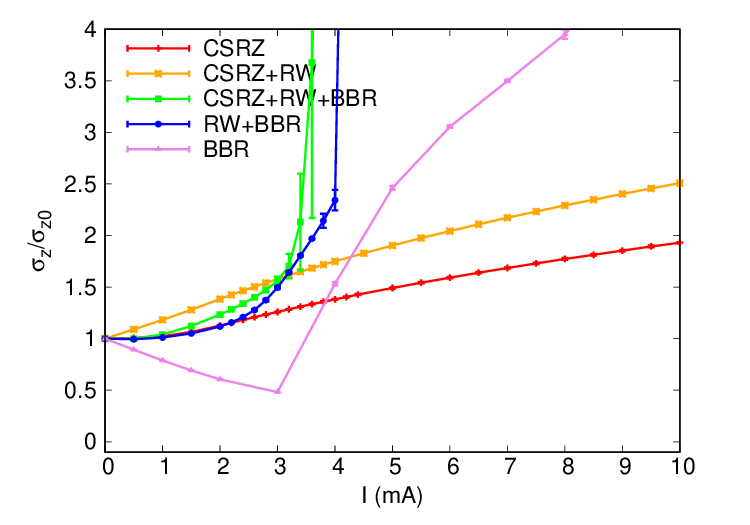}
    \includegraphics[width=\linewidth]{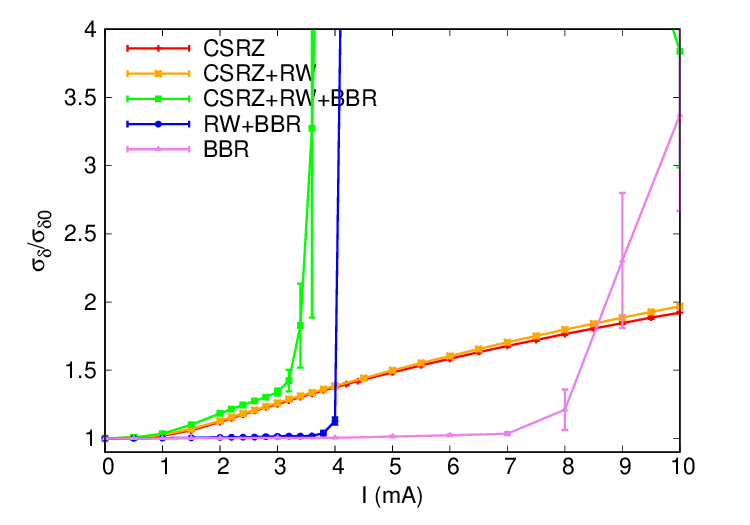}
    \caption{Normalized bunch length and energy spread as a function of bunch current for Elettra 2.0. Simulations were done using Elegant with different combinations of the impedance models presented in Fig.~\ref{fig:ZL-Elettra2.0}. The bunch is with $10^6$ multi-particles and tracking is done for five times of longitudinal damping time. This result is without 3HC effect.}
    \label{fig:BL_ES_Elegant}
\end{figure}

\begin{figure}
    \centering
    \includegraphics[width=\linewidth]{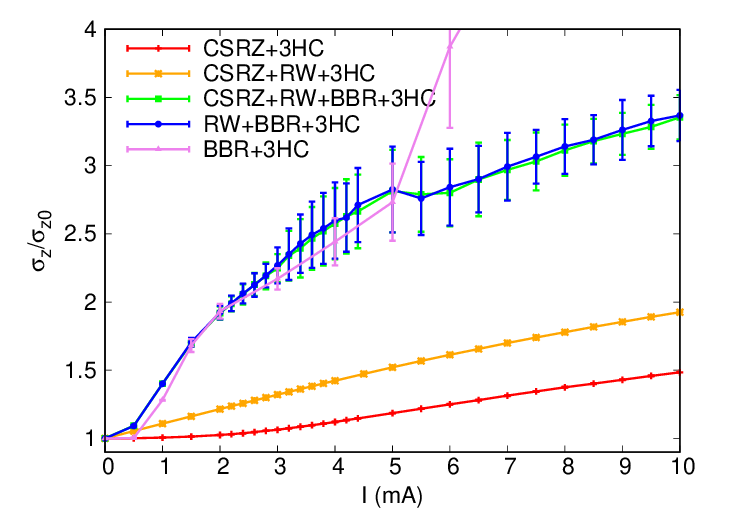}
    \includegraphics[width=\linewidth]{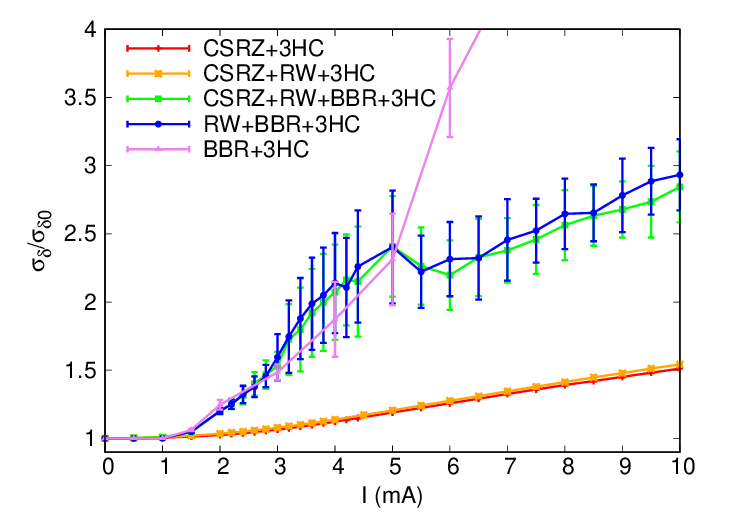}
    \caption{Normalized bunch length and energy spread as a function of bunch current for Elettra 2.0. Simulations were done using Elegant with different combinations of the impedance models presented in Fig.~\ref{fig:ZL-Elettra2.0}. The bunch is with $10^6$ multi-particles and tracking is done for five times of longitudinal damping time. This result is with 3HC effect.}
    \label{fig:BL_Elegant_3HC}
\end{figure}

\begin{figure}
\centering
\includegraphics[width=\linewidth]{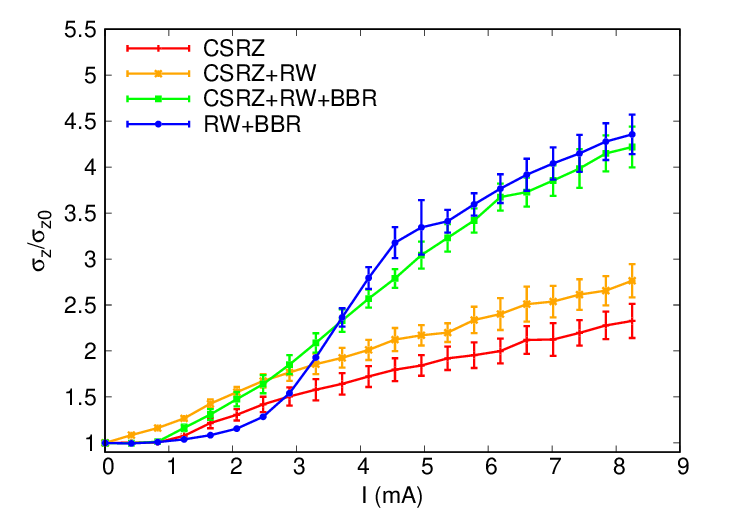}
\includegraphics[width=\linewidth]{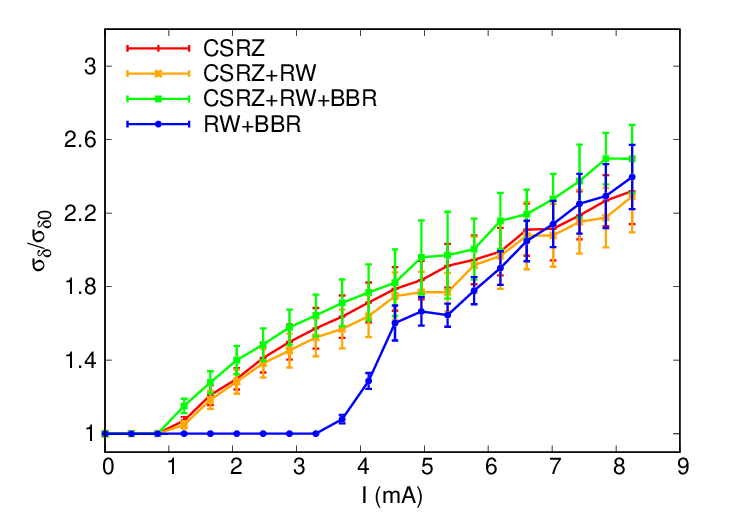}
\caption{Normalized bunch length and energy spread as a function of bunch current for Elettra 2.0. Simulations were done using the VFP solver described in~\cite{bane2010threshold} with different combinations of the impedance models presented in Fig.~\ref{fig:ZL-Elettra2.0}. Initial beam parameters refer to Tab.~\ref{tab:parameters} without 3HC.}
\label{fig:VFP-CSR-RW-BBR-Elettra2.0}
\end{figure}

We explore the interplay between the CSR, RW, and BBR impedances within Elettra 2.0. For Elegant simulations, we track a single bunch comprising $10^{6}$ macroparticles, considering various combinations of these impedances. The CSR impedance used refers to CSRZ data with a cutoff frequency of $k_\text{max}=35$ mm$^{-1}$, without additional filtering of high-frequency impedances. The CSR effects are recognized on top of the RW and BBR impedance effects. Furthermore, this investigation is carried out with and without the 3HC effect. The results of the Elegant and VFP simulations are summarized in Figs.~\ref{fig:BL_ES_Elegant},~\ref{fig:BL_Elegant_3HC}, and~\ref{fig:VFP-CSR-RW-BBR-Elettra2.0}.

In the absence of 3HC, both Elegant and VFP simulations yield nearly identical MWI thresholds for the same impedance models. With the inclusion of the CSR impedance, the threshold current remains close to 1 mA and is unaffected by variations in impedance source combinations. This suggests that the CSR impedance alone determines the MWI threshold when the 3HC is inactive. It is worth noting that, when considering the BBR, Elegant simulations exhibit greater instability above the MWI threshold compared to VFP simulations. This discrepancy may be attributed to the treatment of impedance sources: Elegant simulations lump all sources to a single point per turn, whereas VFP simulations distribute them across approximately 10 points per turn.

In the presence of 3HC, the Elegant simulations reveal that the MWI threshold is set solely by the BBR impedance. As anticipated by the scaling law in Eq.~(\ref{eq:IthCSR}), the CSR effects diminish because of the reduction in peak current resulting from bunch lengthening induced by 3HC. It is intriguing that the MWI threshold decreases when 3HC is applied in conjunction with the BBR impedance. In addition, the observed bunch lengthening appears plausible. This outcome mainly originates from the selection of $f_r=7$ GHz, corresponding to $k_r\sigma_z=0.26$ and $0.66$ for zero-current bunch lengths of 1.8 mm (without 3HC) and 4.5 mm (with 3HC), respectively. These phenomena indicate the sophisticated characteristics of the longitudinal beam dynamics driven by the BBR impedances with $k_r\sigma_z<1$. In the subsequent subsection, we delve into a qualitative discussion of the underlying physics.


%

\subsubsection{Understanding simulation results with the $k_r\sigma_z<1$ BBR model}



\begin{figure}
    \centering
    \includegraphics[width=\linewidth]{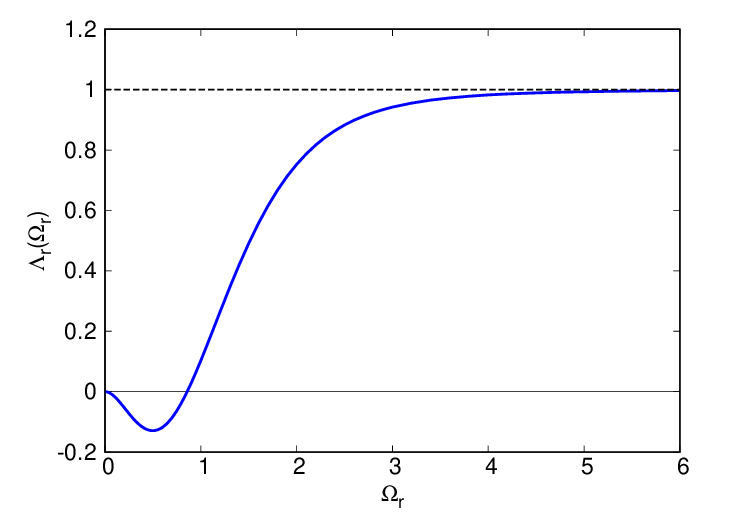}
    \caption{$\Lambda_r(\Omega_r)$ for the effective inductance of a BBR impedance model. The dashed line indicates $\Lambda_r(\Omega_r)$=1 when $\Omega_r\rightarrow \infty$.}
    \label{fig:Leff_over_L0_BBR}
\end{figure}

Using the BBR impedance model from Eq.~(\ref{eq:bbr}), we can calculate the effective inductance experienced by a Gaussian bunch, following the formulations of~\cite{zhou2024theories} (see Table I therein):
\begin{equation}
    L_\text{eff}=L_0 \Lambda_r(\Omega_r),
    \label{eq:LeffBBR1}
\end{equation}
with
\begin{equation}
    \Lambda_r(\Omega_r)=
    -\frac{2\sqrt{\pi}Q\Omega_r}{Q'}
    \left[ \frac{\Omega_rQ'}{\sqrt{\pi}Q}
    + \text{Im} \left[ \Omega_1^2 w(\Omega_1) \right] \right],
\end{equation}
\begin{equation}
    L_0=R_s/(Qk_rc).
\end{equation}
Here, $L_0$ is an inductance calculated from Eq.~(\ref{eq:bbr}) by taking the limit of $k\rightarrow 0$; $Q'=\sqrt{Q^2-1/4}$; $\Omega_r=k_r\sigma_z$; $\Omega_1=\Omega_r(-i/2+Q')/Q$; $w(z)=e^{-z^2}\left[1-i\text{Erfi}(z)\right]$ with $\text{Erfi}(z)$ the imaginary error function. The function $\Lambda_r(\Omega_r)$ with $Q=1$ is plotted in Fig.~\ref{fig:Leff_over_L0_BBR}. The results show that the beam experiences an inductance equivalent to $L_0$, only when $k_r\sigma_z \gg 1$ (that is, $\Lambda_r(\Omega_r)\approx 1$). For $k_r\sigma_z \lesssim 3$, the inductance perceived by the beam is significantly less than $L_0$. In particular, when $k_r\sigma_z < 0.86$, the scaling factor $\Lambda_r(\Omega_r)$ becomes negative, resulting in a deductive impedance for the beam and causing bunch shortening at low currents.

According to the potential-well bunch lengthening theory~\cite{zhou2024theories}, the effective inductance $L_\text{eff}$ denotes the rate of bunch lengthening at low currents. For Elettra 2.0 without the 3HC, $k_r\sigma_z$=0.26, which is well below the critical value of 0.86. This results in $L_\text{eff}<0$ for the chosen BBR model, leading to a bunch shortening, as shown by the purple lines in Fig.~\ref{fig:BL_ES_Elegant}. With the inclusion of the 3HC, the bunch profile adopts a flat top (for example, see~\cite{tavares2014equilibrium}) and significantly deviates from a Gaussian shape. Consequently, Eq.~(\ref{eq:LeffBBR1}) and the theory presented in~\cite{zhou2024theories} are not directly applicable. However, we can still qualitatively conclude that with $k_r\sigma_z = 0.66$, which is close to the critical value, the lengthening induced by the BBR impedance remains marginal, as illustrated by the purple lines in Fig.~\ref{fig:BL_Elegant_3HC}.

The RW impedance exhibits a significant inductive component, as shown in Fig.~\ref{fig:ZL-Elettra2.0}, and prompts bunch lengthening at low currents. Consequently, it compensates for the bunch shortening induced by the BBR impedance in the absence of 3HC, as illustrated by the blue lines in Fig.~\ref{fig:BL_ES_Elegant}. With the 3HC, the RW impedance causes bunch lengthening (see the blue lines of Fig.~\ref{fig:BL_Elegant_3HC}) as expected.

The MWI threshold driven by BBR impedances has been well investigated in the literature~\cite{oide1990longitudinal, mosnier1999microwave, bane2010threshold}. Without harmonic cavities, the MWI threshold as a function of $\Omega_r$ is given by~\cite{oide1990longitudinal}
\begin{equation}
    I_\text{th4}=
    \frac{(E/e)\eta\sigma_\delta^2Q}{R_s} \frac{S_\text{th4}(\Omega_r)}{\Omega_r}.
    \label{eq:IthBBR1}
\end{equation}
According to simulations~\cite{oide1990longitudinal, bane2010threshold}, $S_\text{th4}(\Omega_r)$ is a nonlinear function. Referring to the set of parameters without 3HC in Table~\ref{tab:parameters} and the BBR parameters chosen for Elettra 2.0, Eq.~(\ref{eq:IthBBR1}) predicts an MWI threshold of 4.5 mA with \(S_\text{th4}(\Omega_r) \approx 15\) (see Fig.~1 of~\cite{bane2010threshold}) at \(\Omega_r = 0.26\). This is roughly consistent with the results shown by the purple lines in Fig.~\ref{fig:BL_ES_Elegant}, where a slight increase in energy spread is observed around 4.5 mA, followed by a significant increase around 8 mA. This trend is similar to the observations reported in Figs.~17 and~18 of~\cite{mosnier1999microwave}, where the mechanisms are detailed through instability analysis.

The complexity observed in simulations with the BBR model for Elettra 2.0 arises from the strong potential-well distortion (PWD) when $k_r\sigma_z<1$. The instability mechanism with PWD was first addressed in~\cite{oide1994mechanism}. Investigations~\cite{oide1990longitudinal, bane2010threshold} showed that $S_\text{th4}(\Omega_r)$ in Eq.~(\ref{eq:IthBBR1}) has a minimum around $\Omega_r=0.8$ (it is around 1.2 according to~\cite{mosnier1999microwave}). The beneficial effect of a harmonic cavity is expected when $\Omega_r \gtrsim 1$. For the parameter sets in Table~\ref{tab:parameters}, $\Omega_r$ is only 0.26 without the 3HC and 0.66 with the 3HC. Intuitively, this suggests that the harmonic cavity will lower the MWI threshold, as demonstrated in the simulation results of Fig.~\ref{fig:BL_Elegant_3HC}.

All simulations presented thus far for Elettra 2.0 are comprehensible within the given context. However, it is important to emphasize that the BBR model represents an approximation of the total ring impedance. Moving forward, the plan is to replace the BBR model with a more accurate one developed through detailed bottom-up modeling. This approach will provide a better representation of the ring's impedance characteristics, facilitating more precise and reliable simulation outcomes.

\section{\label{sec:csrSKBLER}CSR effects in SuperKEKB LER}

CSR effects were investigated and found to play a significant role in determining the MWI threshold in SuperKEKB LER~\cite{zhou2014impedance}. In this section, we present a systematic analysis using an updated impedance model for geometric and RW wakes~\cite{ishibashi2024impedance}.

\subsection{\label{sec:impSKBLER} Longitudinal broadband impedance model}

As a collider ring, the primary impedance sources in SuperKEKB LER include RF cavities, resistive walls, collimators, bellows, flanges, clearing electrodes, tapered beam pipe, etc. (see Table 2 of~\cite{ishibashi2024impedance}). The pseudo-Green function wakes of geometric discontinuities were calculated using codes GdfidL~\cite{gdfidl} and ECHO3D~\cite{echo4d} with a driving Gaussian bunch of $\hat{\sigma}$=0.5 mm.
\begin{figure}[hbt!]
\centering
\includegraphics[width=\linewidth]{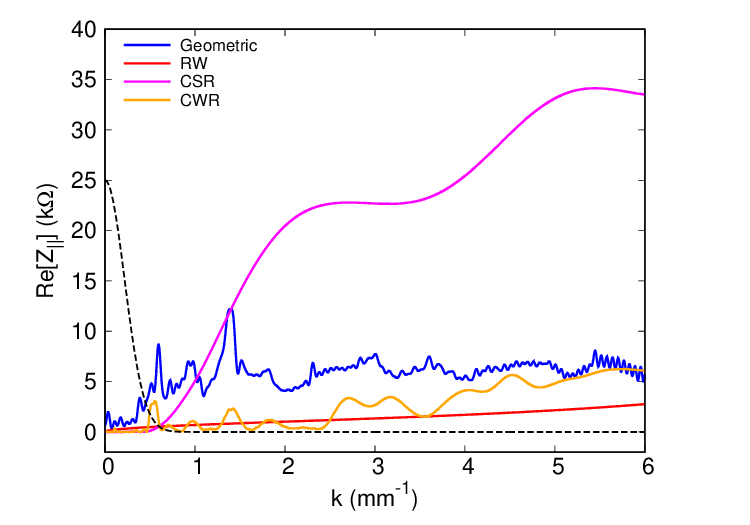}
\includegraphics[width=\linewidth]{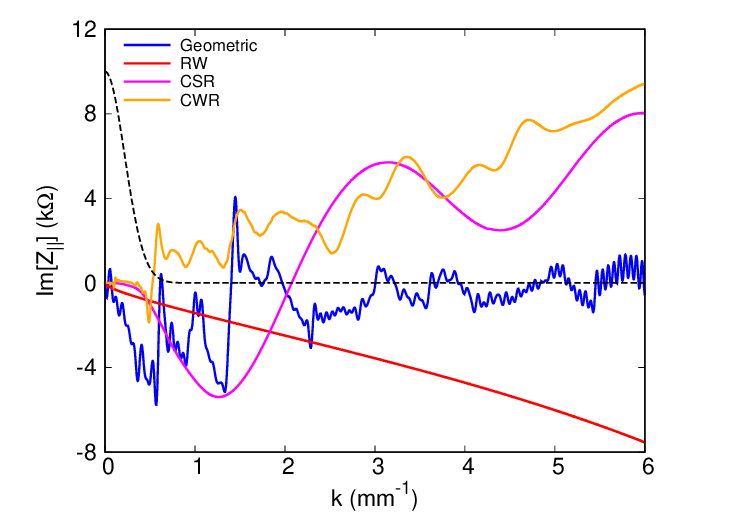}
\caption{Comparison of real and imaginary parts of geometric, RW, CSR, and CWR impedances for SuperKEKB LER. The black dashed lines show the beam spectrum of the nominal Gaussian bunch with $\sigma_{z}=4.6$ mm.}
\label{fig:ZL-superkekB}
\end{figure}

The RW impedance was calculated using the IW2D code. For regular beam pipes, which contains about 90\% of the whole ring, a round aluminum chamber with a radius of 45 mm and a TiN coating of 0.2-$\mu\text{m}$ is used for the impedance calculation.

The CSRZ code is used to calculate the CSR impedance of bending magnets and damping wigglers (so-called coherent wiggler radiation (CWR)). 112 regular bending magnets in the arc sections have 4.2 m length and 74.7 m bending radius. Damping wigglers have a minimum bending radius of around 15 m, a wavelength of 1.1 m, and about 300 periods in total. A square chamber with a full width of 90 mm is used to approximate the boundary condition in the CSR impedance calculation. The total CSR impedance is the sum of a regular bend multiplied by 112 and a superperiod of wigglers (10 periods interleaved with drifts) multiplied by 30.

The four mentioned impedances for SuperKEKB LER are compared in Fig.~\ref{fig:ZL-superkekB}. Given the focus on high-frequency impedances in this study, geometric impedance data were obtained by Fourier transforming the pseudo-Green function wakes, followed by multiplication by $\text{Exp}[k^2\hat{\sigma}^2/2]$ with $\hat{\sigma}=0.5$ mm. This differs from the analyses conducted in~\cite{zhou2024theories}, which focus on potential-well distortion. With a nominal bunch length of $\sigma_z=$4.6 mm, the maximum wavenumber is chosen to be $k_\text{max}$=6 mm$^{-1}$, corresponding to $k_\text{max}\sigma_z$=27.6. This should be sufficient for instability analysis and MWI simulations. It is evident that in the high-frequency region of $k\gg 1/\sigma_z$, the CSR impedance is the most significant source in SuperKEKB LER, although other impedances also contribute notably.

\subsection{\label{sec:anaSKBLER}Instability analysis}

\begin{figure}
\centering
\includegraphics[width=\linewidth]{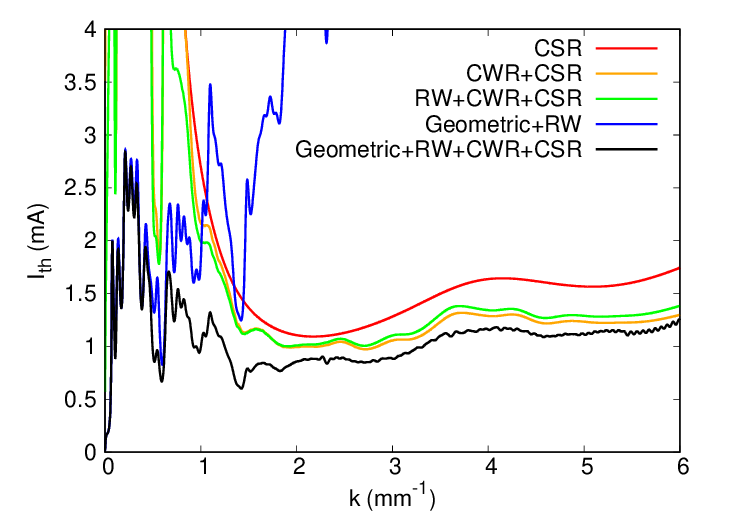}
\caption{Threshold current as a function of wavenumber for SuperKEKB LER with nominal bunch length $\sigma_z$=4.6 mm. The impedance models for the analysis are combinations of CSR, CWR, RW, and geometric impedances as depicted in Fig.~\ref{fig:ZL-superkekB}. Additional beam parameters are given in Tab.~\ref{tab:parameters}.}
\label{fig:Ibth-SuperKEKB-LER}
\end{figure}

Instability analysis refers to the parameter set of SLER with $\sigma_z$=4.6 mm in Table~\ref{tab:parameters}. Considering various combinations of impedance sources shown in Fig.~\ref{fig:ZL-superkekB}, the analysis results are presented in Fig.~\ref{fig:Ibth-SuperKEKB-LER}. These results indicate that high-frequency impedances in the region where $k\sigma_z \gg 1$ set the MWI threshold to around 1 mA and should be properly accounted for in MWI simulations. The CSR instability threshold estimated from the CSRZ data is slightly higher than that estimated from the PP-SS model shown in Table~\ref{tab:parameters}. This discrepancy is mainly due to additional shielding provided by the chamber, which has an aspect ratio of 1 in SuperKEKB LER, significantly different from that of parallel plates.

\begin{figure}
\centering
\includegraphics[width=\linewidth]{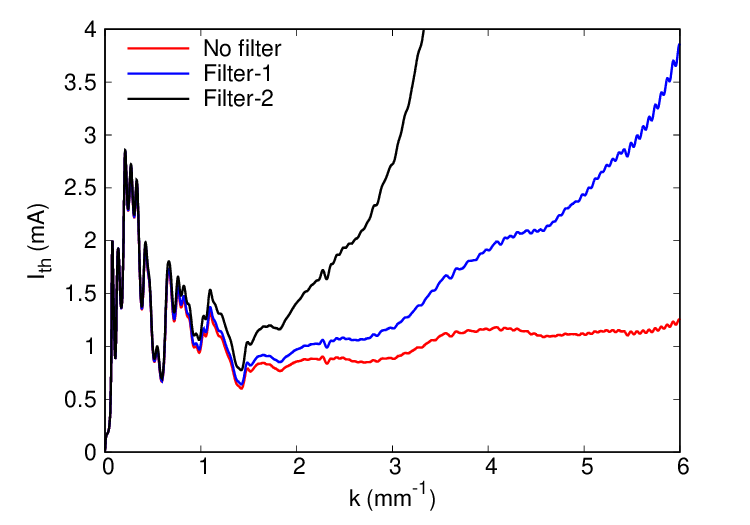}
\caption{Threshold current as a function of wavenumber for SuperKEKB LER with nominal bunch length $\sigma_z$=4.6 mm. The impedance models for the analysis are total impedance (i.e, sum of CSR, CWR, RW and geometric impedance) with no filtering (red line), Gaussian filter $\text{Exp}[-k^2\hat{\sigma}_{1}^2/2]$ (blue line) and Gaussian filter $\text{Exp}[-k^2\hat{\sigma}_{2}^2/2]$ (black line). $(\hat{\sigma}_{1}, \hat{\sigma}_{2})=(0.25, 0.5)$ mm. Additional beam parameters are given in Tab.~\ref{tab:parameters}.}
\label{fig:Ibth-filter-SuperKEKB-LER}
\end{figure}

As noted in the previous section, using pseudo-Green function wakes with a bunch length of $\hat{\sigma}$ in MWI simulations is equivalent to applying a Gaussian filter of $\text{Exp}[-k^2\hat{\sigma}^2/2]$ to the impedance data. Therefore, we also perform an instability analysis with the filtered data to compare with the results without filtering, as shown in Fig.~\ref{fig:Ibth-filter-SuperKEKB-LER}. Once again, we observe that $\hat{\sigma} \lesssim \sigma_z/20$ is preferred to avoid a significant underestimate of the MWI threshold. This will be further supported by simulations presented in the following subsection.

\subsection{\label{sec:simSKBLER}Simulations of microwave instability}

\begin{figure}[hbt!]
\centering
\includegraphics[width=\linewidth]{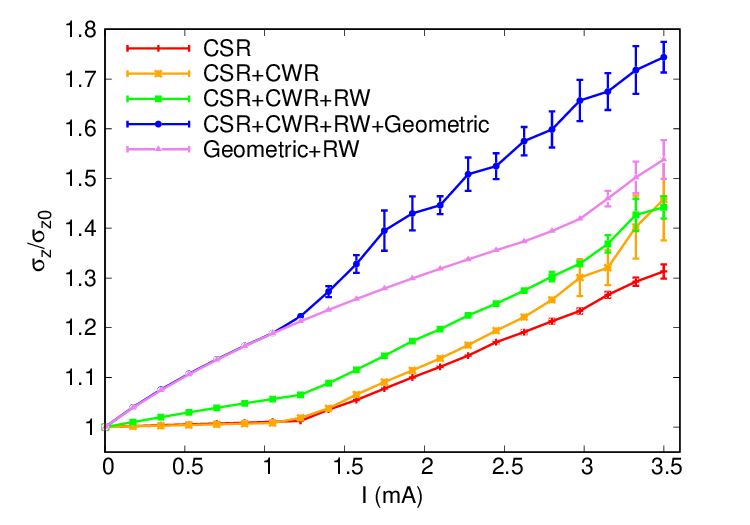}
\includegraphics[width=\linewidth]{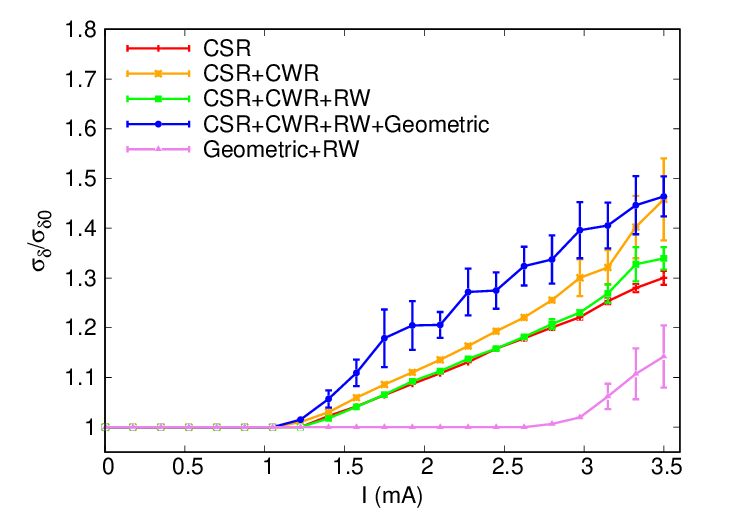}
\caption{Normalized bunch length and energy spread as a function of bunch current for SuperKEKB LER. Simulations were done using the VFP solver described in~\cite{bane2010threshold} with the impedance models of Fig.~\ref{fig:ZL-superkekB}. The initial beam parameters refer to Tab.~\ref{tab:parameters}.}
\label{fig:VFP-SuperKEKB-LER}
\end{figure}

For SuperKEKB LER, we use the VFP solver described in ~\cite{bane2010threshold} for MWI simulations. As the code reads wake data instead of impedance, the Fourier transform is applied to convert the impedance data of Fig.~\ref{fig:ZL-superkekB} into wake data for the simulations. The results with different combinations of wake sources are shown in Fig.~\ref{fig:VFP-SuperKEKB-LER}. It is observed that without inclusion of the CSR impedance, the MWI threshold is around 2.8 mA; when it is included, the threshold reduces to around 1.2 mA, as expected by the scaling laws in Sec.~\ref{sec:instability} (see Table~\ref{tab:parameters}) and the previous instability analysis. Above the MWI threshold, the high-frequency CSR and CWR impedances significantly drive microbunching, leading to an additional increase in bunch length and energy spread. The simulations also indicate that the bunch lengthening driven by regular geometric and RW impedances does not help increase the CSR instability threshold at SuperKEKB LER, as predicted by the scaling law in Eq.~(\ref{eq:IthCSR2}). This discrepancy may be attributed to potential-well distortion.

\begin{figure}[hbt!]
\centering
\includegraphics[width=\linewidth]{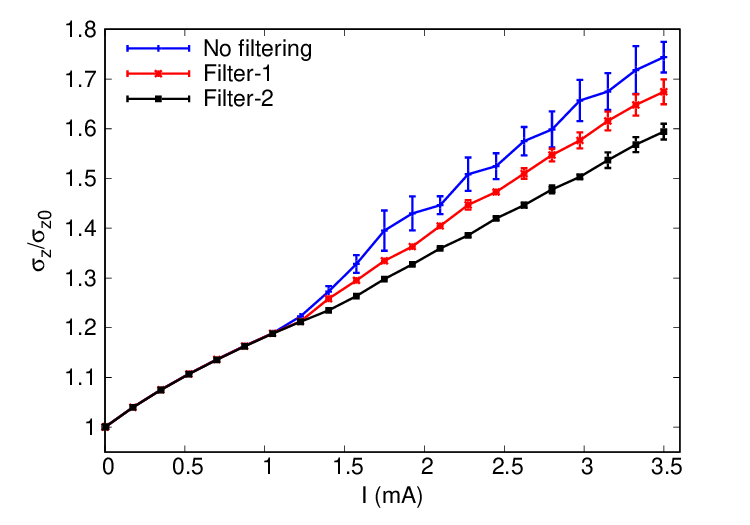}
\includegraphics[width=\linewidth]{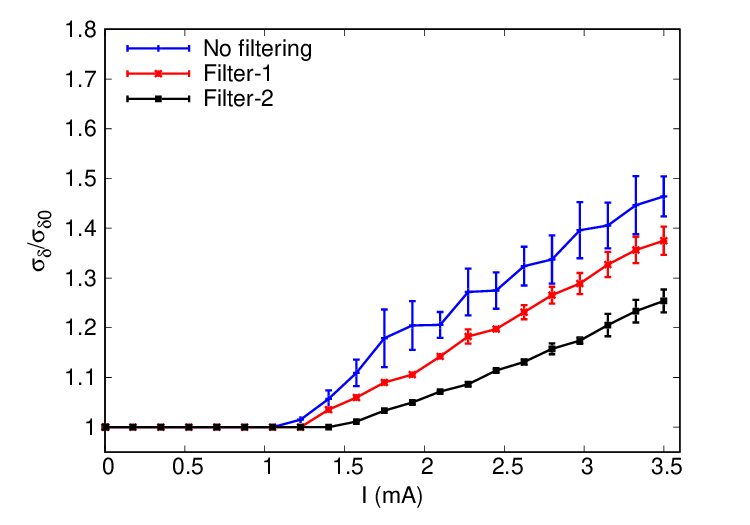}
\caption{Normalized bunch length and energy spread as a function of bunch current for SuperKEKB LER. Simulations were done using the VFP solver described in~\cite{bane2010threshold} with no filtering (blue), Gaussian filter $\text{Exp}[-k^2\hat{\sigma}_{1}^2/2]$ (blue line) and Gaussian filter $\text{Exp}[-k^2\hat{\sigma}_{2}^2/2]$ (black line). $(\hat{\sigma}_{1}, \hat{\sigma}_{2})=(0.25, 0.5)$ mm. Initial beam parameters refer to Tab.~\ref{tab:parameters}.}
\label{fig:VFP-SuperKEKB-LER-Filters}
\end{figure}

Lastly, we examine the effects of filtering the impedance data. The total wakes were calculated from the impedance data in Fig.~\ref{fig:ZL-superkekB} using a filtering function of $\text{Exp}[-k^2\hat{\sigma}^2/2]$ with $\hat{\sigma}=$0, 0.25, and 0.5 mm, respectively. The corresponding results of the MWI simulation are compared in Fig.~\ref{fig:VFP-SuperKEKB-LER-Filters}. Below the MWI threshold, the potential-well distortion caused by low-frequency impedances (typically in the region of $k\sigma_z\lesssim 3$) dominates the bunch lengthening, and the effect of high-frequency impedances (CSR, CWR, and RW) is almost negligible.

Figure~\ref{fig:VFP-SuperKEKB-LER-Filters} also shows that the MWI threshold is clearly affected by the choice of $\hat{\sigma}$, as already observed in the instability analysis of Fig.~\ref{fig:Ibth-filter-SuperKEKB-LER}. Additionally, filtering significantly influences the rate of increase in bunch length and energy spread because the wake forces from microbunching are artificially suppressed by such filtering. Even with $\hat{\sigma}=$ 0 mm, we have chosen an arbitrary cutoff point for the impedance data at $k_\text{max}=$6 mm$^{-1}$. It can be expected that including more data on higher frequency impedances would result in more severe beam instability at bunch current above the MWI threshold. This generally indicates a technical challenge in simulations of unstable beams driven by the CSR impedance. From a physics perspective, impedances up to $k_c$ (critical wavenumber of synchrotron radiation) should be included as fully as possible. From a numerical simulation perspective, the mesh sizes should be as small as possible to accurately capture the impedance information around $k_\text{max}$ for both VFP and macroparticle tracking. Specifically, for VFP and tracking simulations, the phase-space and physical-space regions should be large enough to accommodate the unstable beam. For particle tracking, the number of macroparticles should be increased as the mesh size decreases to avoid numerical noise from the sampling in the construction of the histogram.

\section{\label{sec:summary}Summary}

In this paper, we revisit the theories of microwave instability in low-emittance electron storage rings driven by the CSR impedance in the wavenumber range of $k \gg 1/\sigma_z$. Through instability analysis, we confirm Cai's finding that the CSR instability threshold current is proportional to $(E/e) \eta \sigma_\delta^2 \sigma_z / h$ and is independent of the bending radius of the dipoles. Using the same method, we derive a scaling law of the threshold current for the microwave instability driven by high-frequency resistive-wall impedance. These theories will be useful for quickly checking threshold currents due to CSR and RW impedances, given the machine parameters of an electron storage ring. In addition, the instability analysis helps guide numerical simulations of microwave instability by determining the frequency range of impedances that should be included in the simulations.

Based on the theories developed, the impact of high-frequency CSR and RW impedances is systematically investigated in two electron rings: Elettra 2.0 and SuperKEKB LER, including geometric impedances caused by discontinuities in the vacuum chamber. Using two well-established codes, MWI simulations are performed using different combinations of impedances constructed for these projects. The simulations reveal a complex interplay between various impedances in determining the MWI threshold, but the physics uncovered can be fairly explained by the theories. In particular, the dynamics of MWI can be significantly altered by the inclusion of a harmonic RF cavity, which is widely used in modern fourth-generation ring-based light sources.

The approach developed in this paper, from instability analysis to numerical simulations, is quite general and can be applied to other electron storage rings where high-frequency impedances are a concern in driving microwave instability.

\begin{acknowledgments}
The author, D.Z., thanks K. Ohmi for inspiring discussions on the CSR instability in electron storage rings, Y. Cai for sharing his VFP solver, which was used for the MWI simulations presented in this paper, and G. Stupakov for valuable discussions on CSR theories.
\end{acknowledgments}






\bibliography{csr-ler}

\end{document}